\documentclass[prb, twocolumn, superscriptaddress]{revtex4-2}
\usepackage{amsmath,amssymb,bm}
\usepackage{hyperref}
\usepackage{graphicx}
\usepackage{epstopdf}
\usepackage{latexsym}
\usepackage{braket}
\usepackage{float}
\usepackage{comment}
\usepackage{mathtools}
\usepackage{array}
\usepackage{tabu}
\usepackage{makecell}
\usepackage{multirow}
\usepackage[inline]{enumitem}
\usepackage{cleveref}
\usepackage[usenames,dvipsnames,table,xcdraw]{xcolor}
\usepackage[normalem]{ulem}
\def\BibTeX{{\rm B\kern-.05em{\sc i\kern-.025em b}\kern-.08em
    T\kern-.1667em\lower.7ex\hbox{E}\kern-.125emX}} 

\begin{document}

\title{Elevated Hall Responses as Indicators of Edge Reconstruction}
\author{Sampurna Karmakar}
\affiliation{Indian Institute of Science Education and Research Kolkata, Mohanpur, Nadia - 741246, West Bengal, India}
\author{Amulya Ratnakar}
\affiliation{Aix Marseille Univ, Université de Toulon, CNRS, CPT, Marseille, France}
\affiliation{Indian Institute of Science Education and Research Kolkata, Mohanpur, Nadia - 741246, West Bengal, India}
\author{Sourin Das}
\affiliation{Indian Institute of Science Education and Research Kolkata, Mohanpur, Nadia - 741246, West Bengal, India}
\begin{abstract}
We investigate edge reconstruction scenarios in the $\nu = 1$ quantum Hall state, focusing on configurations with upstream and downstream charge and neutral modes. Our analysis shows that the coexistence of upstream charge and neutral modes in a multiterminal geometry can cause pronounced deviations from the expected quantized values of electrical ($e^2/h$) and thermal ($\pi^2 k_\text{B}^{2}T/3h$) Hall conductance dictated by bulk-boundary correspondence. In particular, we find that both electrical and thermal Hall conductances can be significantly enhanced---exceeding twice their unreconstructed values---offering a clear diagnostic of edge reconstruction.
\end{abstract}
\maketitle
\section{Introduction}
The quantum Hall (QH) state emerges when a two-dimensional electron gas (2DEG) is subjected to a strong perpendicular magnetic field, resulting in a gapped insulating bulk and gapless chiral edge states along the system’s boundaries. According to the bulk-boundary (BB) correspondence principle~\cite{wen1991}, the topological characteristics of the bulk are directly mirrored in the properties of the edge states, resulting in robust and quantized electrical and thermal Hall conductance. This predicted robustness has driven extensive experimental and theoretical research on the electrical transport properties of QH systems over the past several decades. Precision measurements with thermal bias across a Hall bar have become feasible only in the past decade, unlocking exciting opportunities to probe not only charge modes but also neutral modes \cite{Bid2010,Deviatov,Yaron2012,Gurman2012,Venkatachalam2012,Inoue2014} at the QH edge. These advancements have significantly enhanced our ability to measure the thermal Hall conductance~\cite{Etyan_Sourin_2009,Plasmon_scattering2010,Giovani_Sourin_2012,Altimiras2012,Banerjee2017,Banerjee2018,Anindya2021,Anindya2019, Srivastav2022,temp_Enhancement,Melcer2022,melcer2023direct,Melcer2024}.

It is clear that though the BB correspondence had neat predictions for the edge states, the properties of the edge in real QH systems could vary significantly depending on the intricate interplay of disorder, electron-electron interactions, and confining potentials, giving rise to diverse phenomena such as edge reconstruction~\cite{Chklovskii1992,Halperin1993,Wen_1994,Meir_1994,Chklovskii1995,Kun2002,Kun2003,Ganpathy2003,Meir2013,Gefen_2021} and equilibration~\cite{KFP_1994,kane1995,PROTOPOPOV2017287,Park2019,mirlin2019,Park2020,Yuval_incoherent,Yuval_Conductance_plateau,yuval_contacts,Gefen_2022,Maiti_2020,manna2024,manna2023experimentallymotivatedorderlength,Ankur_5by2,pandey2024halfquantizedhallplateausconfined}, etc. It is important to note that these phenomena are not solely determined by the topology of the bulk state but are governed by the specific details of the edge state itself, hence bringing in physics beyond BB correspondence. Also note that this is distinct from other issues related to the correct identification of the bulk state itself. For instance, if we have topologically distinct candidates for the bulk ground states of a given filling fraction $\nu$ (e.g. for $\nu=5/2$~\cite{Tsui_5by2}), then, while the charge Hall coefficient may remain identical for all candidate, the thermal Hall coefficient may set them apart. In general, distinguishing between such candidate states can be challenging. Recent proposals~\cite{Park2020,Yutushi,Ankur_5by2,Park2024} have aimed to address this issue, but it is beyond the scope of this article. 

In this work, we examine the electrical Hall conductance ($G_{H}$) and the thermal Hall conductance ($G^{Q}_{H}$) across a QPC in a six-terminal Hall bar geometry as shown in fig.~\ref{fig:multiterminal}. In absence of the QPC, $G_{H}$ takes quantized integer or fractional values of $e^{2}/h$ (where $e$ is the electron charge and $h$ is Planck's constant) depending on the bulk filling fraction $\nu$ of the QH state, while the thermal Hall conductance, $G^{Q}_{H}$, assumes a quantized value in units of $\pi^2 k_\text{B}^{2}T/(3h)$, where $k_\text{B}$ is the Boltzmann constant and $T$ is the temperature. In this paper, we study the charge and the heat conductance in various edge reconstruction scenarios of the $\nu_B=1$ QH state, leading to the emergence of upstream charge and neutral modes. The charge and heat conductance in the Hall direction attain the universal value dictated by the bulk, independent of transmission through the QPC in the fully equilibrated (incoherent) limit. We show that in the coherent limit, Hall conductances deviate from the value dictated by the BB correspondence and depend on the QPC's transmission. We present that the Hall conductance shows enhancement in the presence of upstream charge and neutral modes as compared to its value for the unreconstructed edge case. The heat conductance shows a very large value if the velocities of the neutral and charge modes are taken to be different, which is in accordance with the experimental findings. This enhanced value of Hall conductance remains larger than that of its unreconstructed counterpart, even if the transmission strength of the QPC is increased.
\section{Electrical and thermal conductance for edge reconstructed quantum Hall system} \label{section_3}
\begin{figure*}[t]
    \centering
    \includegraphics[width=0.98\linewidth]{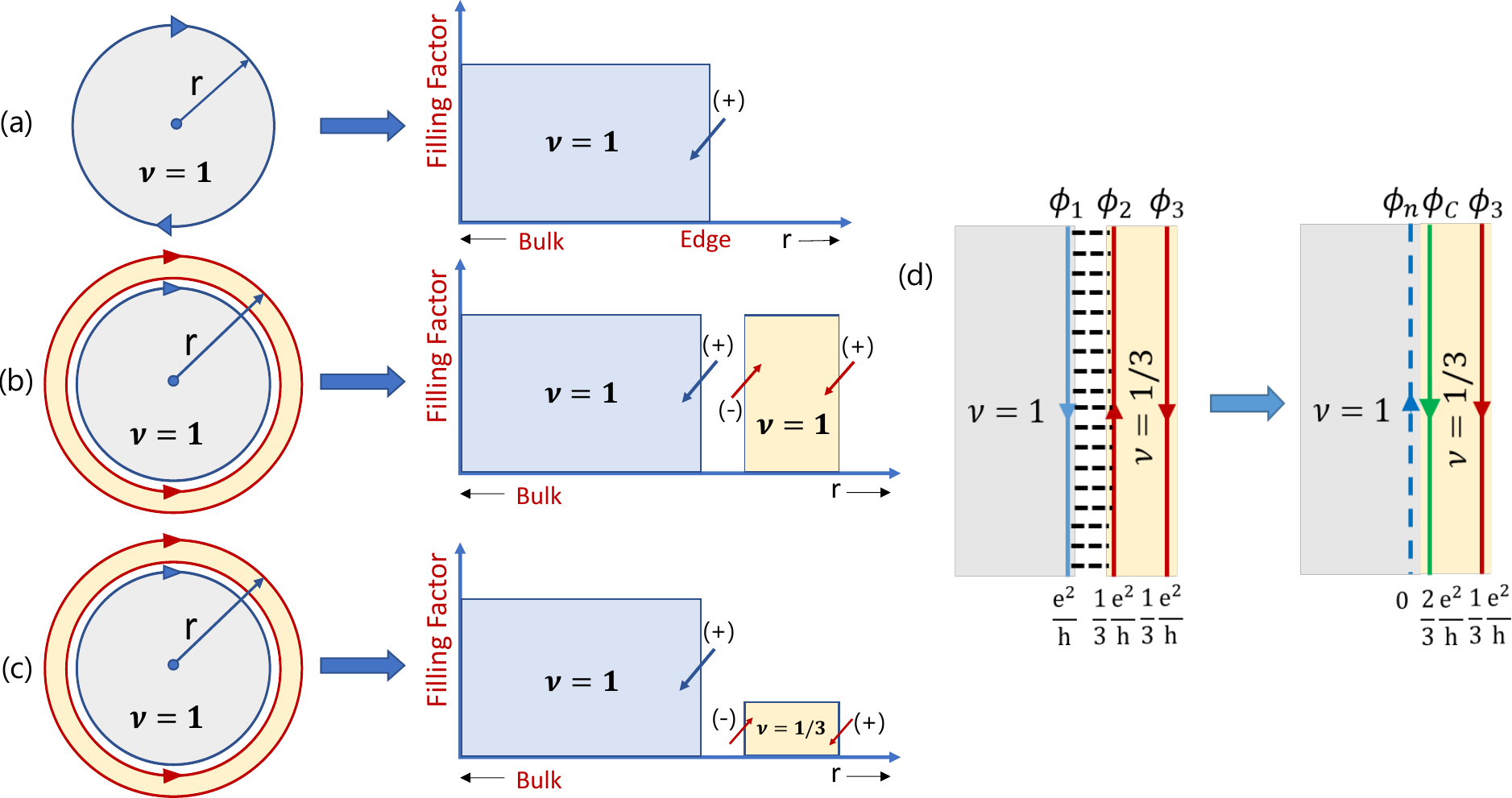}
    \caption{Edge structures of a $\nu=1$ QH system under sharp and smooth confining potential. (a) shows the edge structure and the filling factor, for sharp confinement potential, as a function of distance $r$ from the center of the 2DEG droplet, leading to a single chiral edge at the boundary. (b) and (c) show  the edge structure and the filling factor for smooth confinement potential, as a function of distance $r$ from the center of the 2DEG droplet, leading to edge reconstruction and the deposition of small (b) $\nu = 1$, (c) $\nu = 1/3$ QH bulk at the boundary. This introduces two more chiral (counter-propagating) edges corresponding to $\nu = 1$ and $\nu = 1/3$ QH state respectively. $(+)$ and $(-)$ sign denote downstream and upstream modes respectively. (d) shows further renormalization of the inner two counter-propagating modes of conductance $\nu = 1$ and $\nu = 1/3$, in the presence of impurity-induced backscattering and inter-mode interactions (denoted by dashed black lines) to an edge structure with downstream charge mode of $\nu = 2/3$ conductance (denoted by green line) and an upstream neutral mode (denoted by dashed blue line). The outermost edge of $\nu = 1/3$ remains (denoted by red line) unaffected.}
    \label{fig:Reconstruction}
\end{figure*}
We begin with an introduction to edge reconstruction in QH systems. In these systems, transport along the edges exhibits universal characteristics governed by BB correspondence, although the detailed edge mode structure depends on interelectronic interactions, confining potential, and disorder-induced backscattering along the edge. In a $\nu = 1$ QH system, when the confining potential is sharp and significantly stronger than inter-electron interactions, the edge electron density drops abruptly from its bulk value to zero (see fig.~\ref{fig:Reconstruction}(a)), forming a single chiral Fermi liquid with conductance $e^2/h$. However, for a smoothly varying confining potential over the magnetic length scale $(l_B)$, Chamon and Wen \cite{Wen_1994} modeled the confining potential as a linearly decreasing positive background charge over a length $w$, which leads to the formation of an additional $\nu = 1$ QH strip beyond a critical length scale $w$, introducing two downstream (aligned with the direction of skipping orbits specified by the magnetic field) and one upstream (propagating in the opposite direction of skipping orbits) edge states (see fig.~\ref{fig:Reconstruction}(b)). This edge reconstruction model contradicts several experiments \cite{Moty_2019,Gefen_2016,Paradiso_2012,Pascher_2014,Maiti_2020} for $\nu = 1$ system. A transport experiment \cite{Moty_2019} using a QPC observed a fractional two-terminal conductance plateau of $(1/3)e^2/h$, inconsistent with the integer edge reconstruction. Recent work \cite{Gefen_2021} demonstrated that smoothing the confining potential favors a fractional QH strip at the edge under reconstruction conditions. Increasing potential smoothness stabilizes fractional edge reconstruction, producing a $\nu = 1/3$ QH side strip (see fig.~\ref{fig:Reconstruction}(c)). This structure, with two downstream modes of conductances $(1/3)e^2/h$ and $e^2/h$, and an upstream mode of conductance $(1/3)e^2/h$, is more robust and energetically favorable than integer reconstructions. Interactions and disorder-induced backscattering may further hybridize the inner two counter-propagating charge modes of $\nu = 1$ and $\nu = 1/3$, leading to the Kane-Fischer-Polchinski (KFP) edge structure \cite{KFP_1994} with a downstream charge mode of conductance $(2/3)e^2/h$ and a neutral upstream mode (see fig.~\ref{fig:Reconstruction}(d)).

\begin{figure*}
    \centering
    \includegraphics[width=0.9\linewidth]{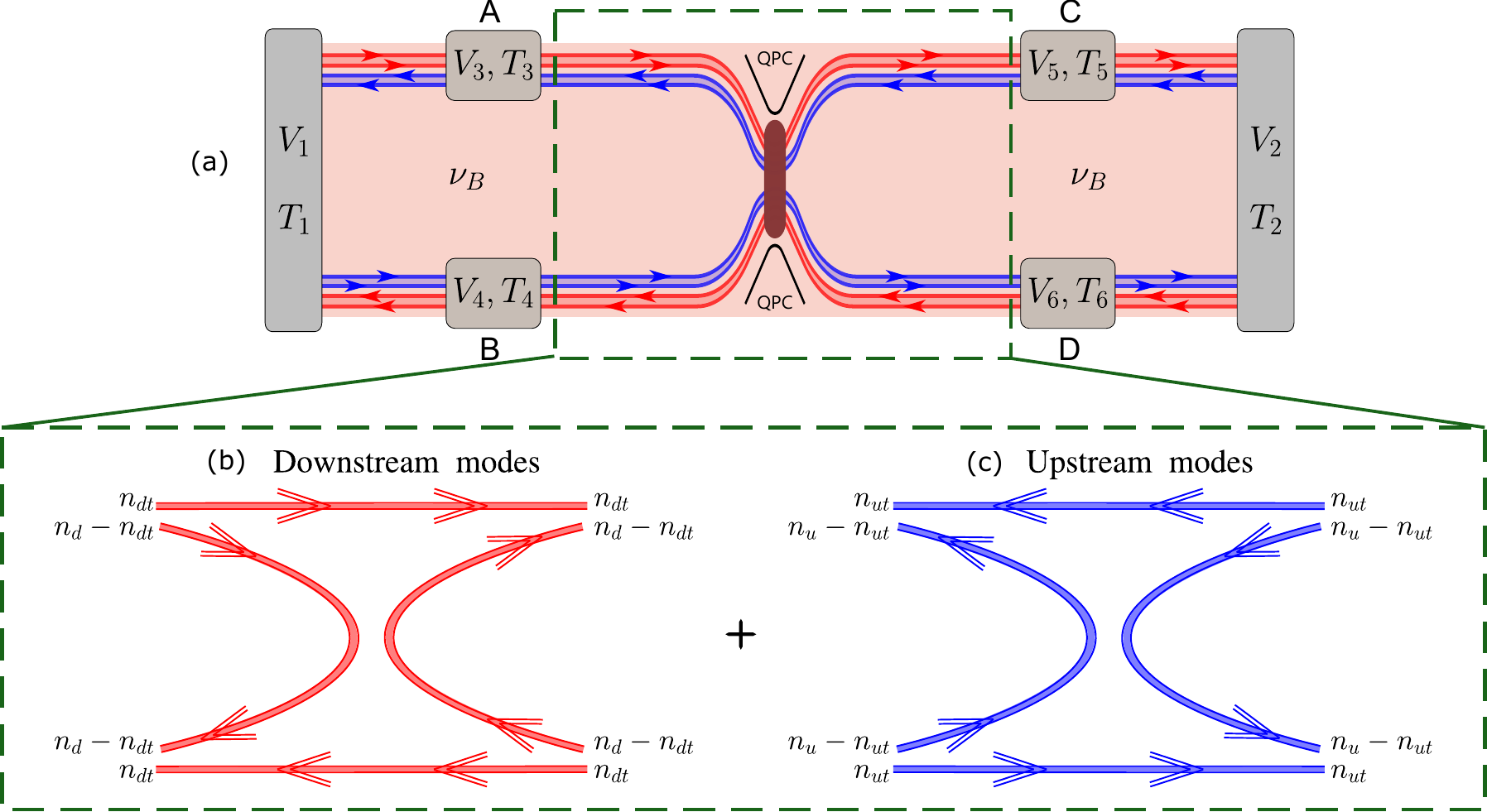}
    \caption{(a) shows the setup to measure the longitudinal and Hall charge, as well as heat conductance in the coherent limit. In this setup, the QH state with filling fraction $\nu_B$ undergoes edge reconstruction, resulting in the emergence of $N$ edge modes, encompassing both downstream (shown in red lines) and upstream (shown in blue lines) modes. The two sources are maintained at voltage and temperature $(V_1, T_1)$ and $(V_2, T_2)$, respectively. The two voltage (temperature) probes at the left [right] of the QPC are maintained at voltage (temperature) $V_3\, (T_3)$ and $V_4 \,(T_4)$ [$V_5\, (T_5)$ and $V_6 \,(T_6)$] respectively.  (b) and (c) shows the charge and heat current splitting for the downstream and upstream modes, respectively. Out of $n_d$ ($n_u$) downstream (upstream) modes, $n_{dt}$ ($n_{ut}$) modes are assumed to be perfectly transmitted at the QPC.}
    \label{fig:multiterminal}
\end{figure*}

With these different QH edge reconstruction scenarios in mind, we examine the effect of edge reconstruction on the charge and heat conductance in the charge non-equilibrated or coherent regime, for a six-terminal setup across a QPC (fig.~\ref{fig:multiterminal}) tuned to a fixed point where the individual edge modes can either perfectly transmit or fully back-reflect. We consider an edge-reconstructed QH system with bulk filling fraction $\nu_B$, with `$N$' edge modes, comprising `$n_u$' and `$n_d$' number of upstream and downstream modes (both consist of $n_{u/d}^c$ and $n_{u/d}^n$ number of charge and neutral modes), respectively, such that $N = n_d + n_u$. We assume that all the charge (neutral) modes have the same velocity, denoted by  $v_c$ $(v_n)$. The filling fraction discontinuity $\nu_{i}$ for the $i^{\text{th}}$ upstream/downstream edge, denoted as $ \nu^{i}_{u/d}$, determines 
\begin{widetext}
    \begin{table*}[t]
\centering
\begin{tabular}{|l|l|}
\hline
\rule{0pt}{3ex}
\hspace{2cm}\textbf{Name} & \hspace{4cm}\textbf{Short Description} 
\rule[-1.5ex]{0pt}{0pt}\\ 

\hline
\rule{0pt}{3ex}
\parbox[t]{6.1cm}{Charge equilibration length ($l^{\text{ch}}_{eq}$)} 
&
\parbox[t]{11.2cm}{Characteristic length scale for the QH system to reach charge equilibration} \\
\hline
\rule{0pt}{3ex}
\parbox[t]{6.1cm}{Thermal equilibration length ($l^{\text{th}}_{eq}$)} 
&
\parbox[t]{11.2cm}{Characteristic length scale for the QH system to reach thermal equilibration} \\
\hline
\rule{0pt}{3ex}
\parbox[t]{6.1cm}{Charge equilibrated (incoherent) but thermal non-equilibrated limit} 
&
\parbox[t]{11.2cm}{$l^{\text{ch}}_{eq}$ $(l^{\text{th}}_{eq})$ is smaller (larger) than the edge length scales $L$ i.e., $l^{\text{ch}}_{eq}<L<l^{\text{th}}_{eq}$} \\
\hline
\rule{0pt}{3ex}
\parbox[t]{6.1cm}{Charge equilibrated (incoherent) but partially thermal equilibrated limit} 
&
\parbox[t]{11.2cm}{Charge equilibrated and weak but finite thermal equilibration ($l^{\text{ch}}_{eq} < l^{\text{th}}_{eq} \lesssim L$)} \\
\hline
\rule{0pt}{3ex}
\parbox[t]{6.1cm}{Full charge equilibrated and full thermal equilibrated (incoherent) limit} 
&
\parbox[t]{11.2cm}{$l^{\text{th}}_{eq}$ is much smaller than the edge length scales ($L >> l^{\text{th}}_{eq}>$ $l^{\text{ch}}_{eq}$ )} \\
\hline
\end{tabular}
\caption{Definitions}
\label{tab:definition}
\end{table*}
\end{widetext}
the charge conductance $ \nu^{i}_{u/d} (e^{2}/h)$ corresponding to the modes (being 0 for neutral modes). Among `$n_d^{c/n}$' downstream charge/neutral modes, let `$n_{dt}^{c/n}$' represent the modes that are perfectly transmitted through the QPC, while `$n_{ut}^{c/n}$' out of the `$n_u^{c/n}$' upstream charge/neutral modes are perfectly transmitted from the QPC. Additionally, let $\nu_{d/u}$ denotes the sum of the filling fraction discontinuities of all downstream/upstream modes, expressed as $\nu_{d/u} = \sum_{i=1}^{n_{d/u}} \nu^{i}_{d/u}$, and $\nu_{dt/ut}$ represents the sum of the filling fraction discontinuities for the downstream/upstream modes that are perfectly transmitted at the QPC, $\nu_{dt/ut} = \sum_{i=1}^{n_{dt/ut}} \nu^{i}_{d/u}$. If $\chi_i$ represents the chirality of the $i^{\text{th}}$ mode, with $\chi_i = +1(-1)$ indicating the downstream (upstream) direction, the edge can exhibit a structure consistent with the constraints imposed by the QH bulk: (i) $\sum_i \chi_i \nu_i = \nu_B$ and (ii) $\sum_i\chi_i =C$, where $C=n_d-n_u$ is the central charge, defined as the difference between downstream and upstream modes. The bulk filling fraction $\nu_{B}$ and the central charge $C$ determine the topologically dictated charge \cite{shtanko2014nonequilibrium} and heat conductance under the fully equilibrated (incoherent) regime, such that the charge Hall conductance and thermal Hall conductance are given by $\nu_{B}(e^{2}/h)$ and $C(\pi^{2}k^{2}_{B}T/3h)$, respectively (see Appendix~\ref{appA} for detailed derivation).  It is well known that in the thermal fully equilibrated limit, the two-terminal heat conductance in the ballistic limit ($G^Q$) is topologically constrained and is proportional to $|n_{d}-n_{u}|$, while the same for the non-equilibrated limit is $(n_{d}+n_{u})$. Most of the experimental studies of thermal transport in the QH regime have been done within the fully equilibrated limit, with only a few experiments in the thermal non-equilibration limit~\cite{Banerjee2017,Banerjee2018,Anindya2021}. It would be interesting to study how thermal conductance, with the reconstructed edge structure, is modified in the presence of finite scattering between the different edge states, which a QPC can achieve.

Consider two sources maintained at voltages (temperatures) $V_1 \,(T_1)$ and $V_2\, (T_2)$ (fig.~\ref{fig:multiterminal}) with $V_{1/2} = V \pm \Delta V/2$ and $T_{1/2} = T \pm \Delta T/2$, respectively, where $ V\,(T)$ denotes the average voltage (temperature) of the system. We consider the linear response regime, where both $\Delta V$ and $\Delta T\longrightarrow 0$. Probes are placed at points A, B, C, and D (fig.~\ref{fig:multiterminal}) to measure the local voltages and temperatures, denoted as  $V_i$ and $T_i$ for $i=3,4,5,6$. These quantities are determined by subjecting to the condition that both the particle number density and the total energy density flowing into each probe match those flowing out. To evaluate these densities, consider a single chiral edge of a QH system with filling fraction $\nu$ and edge velocity $v$, emanating from a source maintained at temperature $T$ and potential $V$.  The corresponding Hamiltonian density is given by
\begin{align}
    \mathcal{H}=\frac{\hbar v}{4\pi\nu}\left( \partial_{x} \phi(x) \right)^{2},
\end{align}
where the bosonic field $\phi(x)$ is expressed as

\begin{align}
    \phi(x)=\nu\bar{\phi_0}+\frac{2\pi}{L}Nx+\sqrt{\frac{2\pi\nu}{L}}\sum_{k>0}\frac{1}{\sqrt{k}}\left(a_ke^{ikx}+a_k^\dagger e^{-ikx}\right),\nonumber
\end{align}
with bosonic creation and annihilation operators obeying $[a_k,a_{k'}^\dagger]=\delta_{kk'}$, and $N$ being the number operator. The thermal expectation $\langle a_{k}^\dagger a_k\rangle=1/(e^{\beta\hbar v k}-1)$, with $\beta=1/(k_\text{B}T)$,  together with the expression for particle number density, $\mathcal{N}=\langle N\rangle/L=\nu \,\frac{eV}{hv}$, allows us to compute the total energy density:
\begin{align}
 \mathcal{E}=\langle \, \mathcal{H} \, \rangle=\frac{1}{v}\left(\nu\,\frac{e^2V^2}{2h}+\frac{\pi^2k^2_\text{B}T^2}{6h}\right).
\end{align}
This energy density consists of two components: the first term represents a purely electrochemical contribution, while the second corresponds to the thermal energy density. Note that the expressions for the charge and heat currents are $I=\nu \frac{e^2}{h}V$ and $J=\frac{\pi^2k^2_\text{B}}{6h}T^2$, respectively. 
In accordance with fig.~\ref{fig:multiterminal}, the particle densities going into the probes are
\begin{align}
\mathcal{N}_{3I} =&\frac{e}{hv_c}\left( \nu_{d} V_{1} + (\nu_{u}- 
   \nu_{ut}) V_{4} + \nu_{ut} V_{5} \right),\nonumber\\
\mathcal{N}_{4I} =&\frac{e}{hv_c}\left(\nu_{u} V_{1} + (\nu_{d}-\nu_{dt}) V_{3} + 
      \nu_{dt} V_{6} \right), \nonumber\\
\mathcal{N}_{5I} =&\frac{e}{hv_c}\left( \nu_{u} V_{2}+(\nu_{d}-\nu_{dt}) V_{6} + 
      \nu_{dt} V_{3}  \right),\nonumber \\
\mathcal{N}_{6I} =&\frac{e}{hv_c}\left( \nu_{d} V_{2} + (\nu_{u}- 
       \nu_{ut}) V_{5} +\nu_{ut} V_{4} \right) ,\nonumber
    \end{align}
and the particle densities coming out from the probes are
\begin{align}
     \mathcal{N}_{iO} =\frac{e}{hv_c}\left(\nu_d+\nu_u\right)V_i\hspace{1cm}\text{ for }i=3,4,5,6.\nonumber
    \end{align}
Similarly, the total energy density going into the probes are
    \begin{widetext}        
    \begin{align}
\mathcal{E}_{3I} =&\frac{e^2}{2hv_c}\left( \nu_{d} V_{1}^2 + (\nu_{u}- 
   \nu_{ut}) V_{4}^2 + \nu_{ut} V_{5}^2 \right)+\frac{\pi^2k^2_\text{B}}{6h}\left(\left(\frac{n_d^c}{v_c}+\frac{n_d^n}{v_n}\right)T_1^2+\left(\frac{n_u^c-n_{ut}^c}{v_c}+\frac{n_u^n-n_{ut}^n}{v_n}\right)T_4^2+ 
  \left(\frac{n_{ut}^c}{v_c}+\frac{n_{ut}^n}{v_n}\right)T_5^2\right),\nonumber\\
  \mathcal{E}_{4I} =&\frac{e^2}{2hv_c}\left(\nu_{u} V_{1} ^2+ (\nu_{d}-\nu_{dt}) V_{3}^2 + 
      \nu_{dt} V_{6}^2 \right)+\frac{\pi^2k^2_\text{B}}{6h}\left(\left(\frac{n_u^c}{v_c}+\frac{n_u^n}{v_n}\right)T_1^2+\left(\frac{n_d^c-n_{dt}^c}{v_c}+\frac{n_d^n-n_{dt}^n}{v_n}\right)T_3^2+ 
  \left(\frac{n_{dt}^c}{v_c}+\frac{n_{dt}^n}{v_n}\right)T_6^2\right), \nonumber\\
  \mathcal{E}_{5I} =&\frac{e^2}{2hv_c}\left( \nu_{u} V_{2}^2+(\nu_{d}-\nu_{dt}) V_{6}^2 + 
      \nu_{dt} V_{3}^2  \right)+\frac{\pi^2k^2_\text{B}}{6h}\left(\left(\frac{n_u^c}{v_c}+\frac{n_u^n}{v_n}\right)T_2^2+\left(\frac{n_d^c-n_{dt}^c}{v_c}+\frac{n_d^n-n_{dt}^n}{v_n}\right)T_6^2+ 
  \left(\frac{n_{dt}^c}{v_c}+\frac{n_{dt}^n}{v_n}\right)T_3^2\right),\nonumber \\
\mathcal{E}_{6I} =&\frac{e^2}{2hv_c}\left( \nu_{d} V_{2}^2 + (\nu_{u}- 
       \nu_{ut}) V_{5}^2 +\nu_{ut} V_{4} ^2\right)+     \frac{\pi^2k^2_\text{B}}{6h}\left(\left(\frac{n_d^c}{v_c}+\frac{n_d^n}{v_n}\right)T_2^2+\left(\frac{n_u^c-n_{ut}^c}{v_c}+\frac{n_u^n-n_{ut}^n}{v_n}\right)T_5^2+ 
  \left(\frac{n_{ut}^c}{v_c}+\frac{n_{ut}^n}{v_n}\right)T_4^2\right) ,\nonumber
    \end{align}
    \end{widetext}
and the energy densities coming out from the probes are
\begin{align}
     \mathcal{E}_{iO} =\frac{e^2V_{i}^2}{2hv_c} \left( \nu_{d} +\nu_{u} \right)+\frac{\pi^2k^2_\text{B}T_i^2}{6h}\left(\frac{n_d^c+n_u^c}{v_c}+\frac{n_d^n+n_u^n}{v_n}\right),\nonumber
    \end{align}  
for $i=3,4,5,6.$ By satisfying the probe conditions, we calculate the net current ($I_{Net}$), Hall voltage ($V_H=V_3-V_4$), {electrical} Hall conductance ($G_H=I_{Net}/V_H$), and longitudinal conductance ($G_L=I_{Net}/\Delta V$) as follows:
\begin{widetext}
\begin{align}
I_{Net} &=\frac{e^2}{h}\frac{2 (\nu_{B}-2\nu_{d})\nu_{dt}\nu_{ut}+(\nu_{B}^2-3\nu_{B}\nu_{d}+3\nu_{d}^2)(\nu_{dt}+\nu_{ut})}{\nu_{B}^2+3\nu_{d}(\nu_d -\nu_{B})-2\nu_{dt}(\nu_B+2\nu_{ut})+2\nu_{d}(\nu_{dt}+\nu_{ut})}\Delta V;\nonumber\\
V_{H} &=\frac{\nu_{d}(\nu_{dt}-\nu_{ut})+\nu_{B}\nu_{ut}}{\nu_{B}^2+3\nu_{d}(\nu_d -\nu_{B})-2\nu_{dt}(\nu_B+2\nu_{ut})+2\nu_{d}(\nu_{dt}+\nu_{ut})}\Delta V;\nonumber\\
V_{36}&=\frac{\nu_{B}^2 +(\nu_{d}-\nu_{B})(3\nu_{d}-2\nu_{ut})}{\nu_{B}^2+3\nu_{d}(\nu_d -\nu_{B})-2\nu_{dt}(\nu_B+2\nu_{ut})+2\nu_{d}(\nu_{dt}+\nu_{ut})}\Delta V;\nonumber\\
G_{H}&=\frac{e^2}{h}\frac{2 (\nu_{B}-2\nu_{d})\nu_{dt}\nu_{ut}+(\nu_{B}^2-3\nu_{B}\nu_{d}+3\nu_{d}^2)(\nu_{dt}+\nu_{ut})}{\nu_{d}(\nu_{dt}-\nu_{ut})+\nu_{B}\nu_{ut}};\nonumber\\
G_{L}&=\frac{e^2}{h}\frac{2 (\nu_{B}-2\nu_{d})\nu_{dt}\nu_{ut}+(\nu_{B}^2-3\nu_{B}\nu_{d}+3\nu_{d}^2)(\nu_{dt}+\nu_{ut})}{\nu_{B}^2+3\nu_{d}(\nu_d -\nu_{B})-2\nu_{dt}(\nu_B+2\nu_{ut})+2\nu_{d}(\nu_{dt}+\nu_{ut})};\nonumber\\
G_{36} &=\frac{e^2}{h}\frac{2 (\nu_{B}-2\nu_{d})\nu_{dt}\nu_{ut}+(\nu_{B}^2-3\nu_{B}\nu_{d}+3\nu_{d}^2)(\nu_{dt}+\nu_{ut})}{\nu_{B}^2 +(\nu_{d}-\nu_{B})(3\nu_{d}-2\nu_{ut})}.
\label{Eq:electrical_conductances}
\end{align}    
\begin{table*}[t]
\centering
\resizebox{0.98\textwidth}{!}{  
\begin{tabular}{||c|c|c|c|c|c|c|c|c|c|c|c||} 
 \hline
\multicolumn{1}{||c|}{} & 
\multicolumn{1}{c|}{Reconstructed edge} & 
\multicolumn{2}{c|}{Transmitted modes} & 
\multicolumn{1}{c|}{$G_{L}$} & 
\multicolumn{1}{c|}{$G_{36}$} & 
\multicolumn{1}{c|}{$G_{H}$} & 
\multicolumn{1}{c|}{$G$} & 
\multicolumn{1}{c|}{$G^{Q}_{L}$} & 
\multicolumn{1}{c|}{$G^{Q}_{36}$} & \multicolumn{1}{c|}{$G^{Q}_{H}$} & 
\multicolumn{1}{c||}{ $G^Q$} \\ [0.5ex] 
 \hline\hline
 
\multirow{3}{*}{1.} & \multirow{3}{*}{+1, -1, +1} 
& +1, -1, +1 & Ballistic & 9/7 & 9/5 & 3 & 3 & 9/7 & 9/5 & 3 &3 \\ \cline{3-12}

& & +1, -1 & \multirow{2}{*}{with QPC} & 8/9 & 8/5 & \cellcolor{pink!40}{\textbf{8}} & 2 & 8/9 & 8/5 & \cellcolor{pink!40}{\textbf{8}} &2\\ \cline{3-3} \cline{5-12}

& & +1 & & 7/9 & 1 & \cellcolor{pink!40}{\textbf{7/2}} & 1 & 7/9 & 1 & \cellcolor{pink!40}{\textbf{7/2}}&1 \\
\hline\hline

\multirow{3}{*}{2.} & \multirow{3}{*}{+1/3, -1/3, +1} 
& +1/3, -1/3, +1 & Ballistic & 65/63 & 65/57 & 13/9 & 5/3 & 9/7 & 9/5 & 3&3 \\ \cline{3-12}

& & +1/3, -1/3 & \multirow{2}{*}{with QPC} & 32/81 & 32/57 & \cellcolor{pink!40}{\textbf{32/9}} & 2/3 & 8/9 & 8/5 & \cellcolor{pink!40}{\textbf{8}} &2\\ \cline{3-3} \cline{5-12}

& & +1 & & 7/23 & 1/3 & \cellcolor{pink!40}{\textbf{7/4}} & 1/3 & 7/9 & 1 & \cellcolor{pink!40}{\textbf{7/2}} &1\\
\hline\hline

\multirow{2}{*}{3.} & \multirow{2}{*}{+1/3, 0, +2/3} 
& +1/3, 0, +2/3 & Ballistic & 1 & 1 & 1 & 1 & $\frac{8+\Tilde{v}^2}{(\Tilde{v}+1)^2+3}$ & $\frac{8+\Tilde{v}^2}{5 -(\Tilde{v}-1)^2}$ & $\frac{8+\Tilde{v}^2}{4 -\Tilde{v}^2}$&3 \\ \cline{3-12}

& & +1/3 & with QPC & 1/3 & 1/3 & \cellcolor{pink!40}{\textbf{1}} & 1/3 & $\frac{(4+3\Tilde{v})}{(\Tilde{v}+2)^2}$ & $\frac{(4+3\Tilde{v})}{(\Tilde{v}+1)^2+3}$ & \cellcolor{blue!20}{\textbf{$2+\frac{3 \Tilde{v}}{2}$}} &1\\
\hline  

\end{tabular}
}
\caption{shows both the transverse and longitudinal electrical (in units of $e^2/h$) and thermal conductance  (in units of $(\pi^2 k_\text{B}^2T/3h$) for a QH state with a bulk filling fraction $\nu_B=1$ and here $\Tilde{v}=v_c/v_n$ is the ratio of the velocities of charge and neutral modes. Various scenarios of QH edge reconstruction, including both integer (case 1 as in fig.~\ref{fig:Reconstruction} (b) with $\nu=1$ reconstructed strip) and fractional (case 2 as in fig.~\ref{fig:Reconstruction} (c) with $\nu=1/3$ reconstructed strip and case 3 as in fig.~\ref{fig:Reconstruction} (d) with further reconstruction of inner downstream $\nu=1$ and upstream $\nu=1/3$ mode) edge reconstructions are considered and represented with different filling fraction discontinuities for the reconstructed modes written from left to right in the outward edge towards the bulk. The downstream and upstream modes are denoted by `+' and `-' signs, respectively. Different cases are considered for the perfectly transmitting edge modes at the QPC along with the ballistic case. $G_L$ $(G_L^Q)$, $G_H$ $(G_H^Q)$ and $G$ $(G^Q)$ denotes the electrical (thermal) longitudinal, Hall and two terminal conductance, respectively. $G_{36}$ $(G_{36}^Q)$ is the electrical (thermal) conductance measured between the probe $V_3$ and $V_6$ ($T_3$ and $T_6$). The possibility for charge (thermal) Hall conductance values to exceed the corresponding two-terminal counterpart $G$ ($G^{Q}$) is indicated by pink and blue colored blocks.}
\label{tab:Table1}
\end{table*}
\end{widetext}
Here, we have employed the bulk constraint: $(\nu_d-\nu_u)=\nu_B.$ Additionally, we include the expressions for $V_{36}$ and $G_{36}$, which represent the voltage difference and charge conductance measured between the voltage probes $V_3$ and $V_6$, respectively. The two-terminal charge conductance, $G$, is equal to $(\nu_{dt}+\nu_{ut})(e^2/h)$. From eq.~(\ref{Eq:electrical_conductances}), it follows that, in the absence of upstream modes in the QH system, i.e., when $\nu_u=0$ (and consequently $\nu_{ut}=0$), the Hall conductance $\nu_{B}(e^2/h)$ becomes equal to the longitudinal conductance in the ballistic limit ($\nu_{dt}=\nu_d$). 

After applying the probe conditions, similar to the charge transport calculation, we compute the net thermal current $J_{Net}$ flowing from left to right, a transverse temperature difference $(T_{H} = T_{3}-T_{4})$ and this leads to the determination of thermal Hall conductance ($G^{Q}_{H}=J_{Net}/T_H$) and longitudinal conductance ($G^{Q}_{L}=J_{Net}/\Delta T$).

For an unreconstructed $\nu_B=1$ QH system with a fully transmitting QPC, the Hall conductance is equal to the longitudinal conductance, $G_H = G_L = e^2/h$. In the presence of finite backscattering ($t<1$), $G_{L} < G_{H}$, where $G_{L} = te^{2}/h$. For the case of $\nu_B=N$, with a fully transmitting QPC, $G_{L}=G_{H} = Ne^{2}/h$~\cite{AHM1987}, even when the Fermi velocities of the edge modes are different. An analogous analysis shows that the thermal conductance likewise satisfies $G_{L}^Q=G_{H}^Q = N\pi^2 k_\text{B}^{2}T/(3h)$. Here, we emphasize that, if we consider the QH system under edge reconstruction, leading to the emergence of upstream edge modes, the charge and the heat Hall conductance can attain a value larger than the unreconstructed QH state counterpart in the coherent limit. This enhancement in conductance is highlighted and shown in Table~\ref{tab:Table1}, which presents the charge and heat conductance in the longitudinal and Hall directions, considering different scenarios of edge reconstructions of a $\nu_B = 1$ QH bulk system. As expected, in the presence of upstream modes, the quantization of charge and thermal Hall conductance is modified from the values dictated by the bulk, even for a fully transmitting case (QPC is fully open) in the coherent limit. We point out that while non-integer thermal conductance has recently been shown to arise from full edge-mode equilibration at engineered interfaces in Abelian systems without invoking Majorana physics~\cite{roy2025}, our work demonstrates that such fractional values of thermal conductance can also naturally emerge in the strictly coherent limit. For a $\nu_B=1$ bulk, these non-integer values are driven entirely by edge reconstruction, specifically the coexistence of upstream and downstream modes and their subsequent mixing at the measurement probes.  Crucially, the resulting conductances are inherently sensitive to the chosen measurement geometry; for instance, multi-terminal conductances (e.g., $G_L$, $G_L^Q$ or $G_{36}$, $G_{36}^Q$) generally differ from their two-terminal counterparts ($G$, $G^Q$).  Furthermore, the charge and the heat conductance are enhanced when the QH edge undergoes reconstruction. This enhancement in conductance increases further if we tune the QPC transparency (highlighted in pink in Table~\ref{tab:Table1}). We also show that, if the velocity of the neutral modes ($v_{n}$) and the charge modes are taken to be different $(v_{n}<v_{c}$ with $v_{c}/v_{n}\sim 10$), as experimentally observed~\cite{Bocquillon2013,Bid2010,Neutral_vel2,Feldman}, the thermal Hall conductance can go to a very large value, as can be seen from the highlighted value in blue. It is important to note that this amplification in Hall conductance (charge and heat) relies directly on the presence of upstream modes.
\section{Conclusion}\label{section_4}
This study investigates electrical and thermal transport in QH systems with a QPC in the coherent limit. In the absence of edge reconstruction, where all edge modes are co-propagating, the Hall conductance exhibits a quantized value dictated by the BB correspondence. However, this behavior changes in the presence of edge reconstruction. When the reconstructed edges support both downstream and upstream charge and neutral modes, and the QPC is tuned to a conductance plateau (such that each mode is either fully transmitted or fully reflected), the Hall conductance can deviate significantly from its topologically expected value. Notably, these conductance values can exceed those of the unreconstructed edge in the coherent limit.

In contrast, edge equilibration in the incoherent limit tends to restore the universality lost in the coherent regime. Full charge equilibration of the edge modes recovers the quantized electrical Hall conductance in accordance with the BB correspondence. Nevertheless, because the thermal equilibration length is typically much larger than the charge equilibration length~\cite{Anindya2021}, the edges can remain thermally non-equilibrated over experimentally relevant length scales. This leads to deviations of the thermal Hall conductance from its topologically dictated value. Once the system length exceeds the thermal equilibration length, the thermal Hall conductance regains its universal quantization, thereby reflecting the underlying bulk topology (see Appendix~\ref{appA}). To probe this topology beyond conductance measurements, we further investigate excess shot noise, which offers deeper insight into the role of upstream neutral modes and equilibration processes. Motivated by earlier studies~\cite{Bunched2003,bid2009,Gefen_2021,Biswas2022,Sabo2017,MannaNoise2024} suggesting that Fano factor encodes information about the bulk filling fraction at low temperatures, we compute the shot noise for a $\nu=1$ QH system in the incoherent regime. We demonstrate that the resulting Fano factor exactly matches the bulk filling fraction (Appendix~\ref{appB}).

Our results open several directions for future research, including the exploration of partial equilibration regimes and extensions to other filling fractions, especially non-Abelian states, with a focus on thermal transport. Such studies may provide systematic probes of edge reconstruction, enable clear distinctions between different equilibration regimes, and identify constituent edge modes. Since our proposal relies on standard conductance and noise measurements, its experimental implementation should be readily accessible.
\begin{acknowledgments}
S.D. would like to thank Philip Kim for insightful discussions on ideas presented in this work during the meeting (code: ICTS/qm100/2025/01) at International Centre for Theoretical Sciences (ICTS). It is a pleasure to thank Ankur Das for collaboration during the initial stages. The authors thank Sourav Manna and Yuval Gefen for fruitful discussions and clarifications regarding their work on edge equilibration during the Workshop on ``Quantum Systems in Low Dimensions" and the Conference on ``Quantum Matter in Low Dimensions: Quantum Transport, Entanglement, and Beyond" held at IIT Gandhinagar. S.K. acknowleges  support from the Prime Minister's Research Fellowship (PMRF) scheme  of  the Ministry of Education, Government of India (PMRF ID: 0501977). A.R. is supported by “ANYHALL” (Grant ANR No. ANR-21-CE30-0064-03). A.R. also received support from the French government under the France 2030 investment plan, as part of the Initiative d’Excellence d’Aix-Marseille Université—A*MIDEX. This research (S.D.) was supported in part by the International Centre for Theoretical Sciences (ICTS) for the Discussion Meeting on ``A Hundred Years of Quantum Mechanics" (code: ICTS/qm100/2025/01).
\end{acknowledgments}
\appendix
\renewcommand{\thefigure}{\Alph{section}}
\section{Hall conductance at full charge and thermal equilibration}\label{appA}
\begin{figure}[t]
    \centering
    \includegraphics[width=0.98\linewidth]{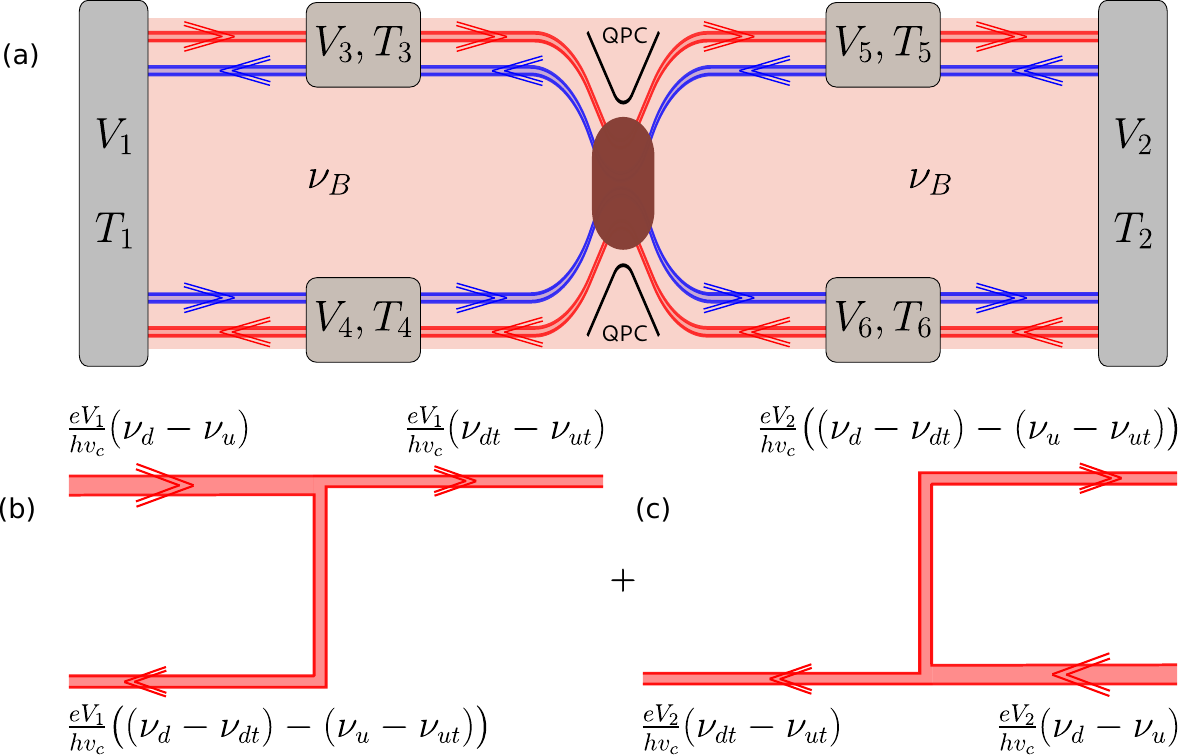}
    \caption{(a) set up for calculating electrical and thermal conductance for a system of bulk filling fraction $\nu_B$, featuring edge reconstruction with both downstream (red lines) and upstream (blue lines) modes. (b) shows the reorganization of the edges into new effective modes after charge equilibration.}
    \label{fig:equilibrate}
\end{figure}

Motivated by the finding in Ref.~[\onlinecite{Anindya2021}] that the charge equilibration length is typically much shorter than the thermal equilibration length, we first consider the regime of full charge equilibration at the edge (with no equilibration at the QPC, since its length scale is much smaller than the charge equilibration length), while thermal equilibration is not assumed. We consider all the system length scales to be greater than the charge equilibration length. Once the charge equilibration is established, all the edge modes propagate with the same equilibrated potential. The net conductance is then solely determined by the net chirality of the edge set by the magnetic field (see fig.~\ref{fig:equilibrate}(b)), washing out the individual contributions as in the coherent regime.

Similar to the set up of fig.~\ref{fig:multiterminal}, we assume a QH system with bulk filling fraction $\nu_B$ (see fig.~\ref{fig:equilibrate}(a)), with $n_d$ downstream and $n_u$ upstream modes, each comprising $n_{u/d}^c$ charge and $n_{u/d}^n$ neutral modes. All the charge (neutral) modes are considered to have same velocity $v_c(v_n)$. Of the $n_d^{c/n}$ downstream charge/neutral modes, $n_{dt}^{c/n}$ are perfectly transmitted through the QPC, while $n_{ut}^{c/n}$ of the $n_u^{c/n}$ upstream modes are perfectly transmitted across it. The sum of the filling fraction discontinuities of all downstream/upstream modes is $\nu_{d/u}$, and $\nu_{dt/ut}$ is for the modes perfectly transmitted at the QPC.

The two sources are at voltages (temperatures) $V_1 (T_1)$ and $V_2 (T_2)$ with $V_{1/2} = V \pm \Delta V/2$ and $T_{1/2} = T \pm \Delta T/2$, respectively, where $ V(T)$ is the average voltage (temperature) and both $\Delta V$ and $\Delta T\rightarrow 0$. We denote the equilibrated voltage at the top left (bottom right) and the bottom left (top right) as $V_{3(6)}$ and $V_{4(5)}$, respectively, in the six terminal geometry (fig.~\ref{fig:equilibrate}(a)). Since we consider $\nu_d > \nu_u$, incoherent equilibration among all downstream and upstream edge modes results in a common local electrochemical potential which is equal to the source potential $V_{1/2}$~\cite{Yuval_incoherent}. Imposing the condition of a vanishing net particle number density at the top left (bottom right) voltage probe leads to
\begin{equation}
    \frac{e}{hv_c}(\nu_d-\nu_u)V_{1(2)}=\frac{e}{hv_c}(\nu_d-\nu_u)V_{3(6)},\nonumber
\end{equation}
giving $V_{3(6)}=V_{1(2)}$. Similarly, at the bottom left (upper right) probe, the condition is
\begin{align}
\frac{e}{hv_c}(\nu_d-\nu_u)V_{4(5)}= \frac{e}{hv_c}\Big(\big((&\nu_d-\nu_u)-(\nu_{dt}-\nu_{ut})\big)V_{1(2)}\nonumber\\ 
     &+(\nu_{dt}-\nu_{ut})V_{2(1)}\Big),
\end{align}
which simplifies to
\begin{align}
    V_{4(5)}=&V_{1(2)}\mp (V_1-V_2)\frac{\nu_{dt}-\nu_{ut}}{\nu_d-\nu_u}.
\end{align}
Since the net charge current flowing from left to right is
\begin{equation}
    I_{net}=\frac{e^2}{h}(\nu_d-\nu_u)(V_1-V_4),
\end{equation}
the electrical Hall conductance becomes to be proportional to bulk filling fraction:
\begin{equation}
    G_H=\frac{I_{net}}{V_H}=\frac{I_{net}}{V_3-V_4}=\frac{e^2}{h}(\nu_d-\nu_u)=\frac{e^2}{h}\nu_B.
\end{equation}
Note that while the full charge equilibration restores the quantized electrical Hall conductance in accordance with BB correspondence, the edges can remain thermally non-equilibrated due to larger equilibration length; consequently, the thermal Hall conductance deviates from its topologically dictated value.\\
In a similar manner, we next compute the heat conductance assuming thermal equilibration, which in turn implies charge equilibration, since the charge equilibration length is much shorter than the thermal equilibration length. We consider all the system length scales involved to be greater than the thermal equilibration length. Under these conditions, all edge modes attain same temperature over this length scale. Given $n_d > n_u$, the temperature of the edges equilibrates to that of the downstream source~\cite{Yuval_incoherent}. Hence following the same probe condition that total energy density entering and leaving the probe is equal, at top left (bottom right) probe, 
\begin{align}
  & \frac{e^2V_{1(2)}^2}{2h} \frac{(\nu_d-\nu_u)}{v_c}+\frac{\pi^2k_B^2T_{1(2)}^2}{6h}\left(\frac{n_d^c-n_u^c}{v_c}+\frac{n_d^n-n_u^n}{v_n}\right)\nonumber\\
   &= \frac{e^2V_{3(6)}^2}{2h} \frac{(\nu_d-\nu_u)}{v_c}+\frac{\pi^2k_B^2T_{3(6)}^2}{6h}\left(\frac{n_d^c-n_u^c}{v_c}+\frac{n_d^n-n_u^n}{v_n}\right).\nonumber
\end{align}
Since we already have $V_{3(6)}=V_{1(2)}$, this directly yields $T_{3(6)}=T_{1(2)}$. Next for the bottom left probe,
\begin{align}
   \frac{e^2V_{4}^2}{2hv_c} &(\nu_d-\nu_u)+\frac{\pi^2k_B^2T_{4}^2}{6h}\left(\frac{n_d^c-n_u^c}{v_c}+\frac{n_d^n-n_u^n}{v_n}\right)\nonumber\\
   &= \frac{e^2}{2hv_c}\left(V_{1}^2 (\nu_d-\nu_u) + (V_2^2-V_{1}^2)(\nu_{dt}-\nu_{ut})\right)\nonumber\\
   &\quad+\frac{\pi^2k_B^2}{6h}\bigg((T_2^2-T_1^2) \left(\frac{n_{dt}^c-n_{ut}^c}{v_c}+\frac{n_{dt}^n-n_{ut}^n}{v_n}\right) \nonumber\\
   &\quad\quad\quad+T_1^2 \left(\frac{n_d^c-n_u^c}{v_c}+\frac{n_d^n-n_u^n}{v_n}\right)\bigg).
\end{align}
Solving this in linear response regime gives
\begin{align}
    T_4=T_1-\Delta T X_t/X
\end{align}
where we have defined $X_t=\left(\frac{n_{dt}^c-n_{ut}^c}{v_c}+\frac{n_{dt}^n-n_{ut}^n}{v_n}\right)$ and $X=\left(\frac{n_d^c-n_u^c}{v_c}+\frac{n_d^n-n_u^n}{v_n}\right)$.  The net heat current flowing from left to right is then given by
\begin{align}
    J_{net}=(n_d-n_u)\frac{\pi^2k_B^2(T_1^2-T_4^2)}{6h}=C\frac{\pi^2k_B^2T\Delta T}{3h} \frac{X_t}{X},
\end{align}
where $C=n_d-n_u$ is the central charge. Consequently, thermal Hall conductance reads 
\begin{align}
    G_H^Q=\frac{J_{net}}{T_H}=\frac{J_{net}}{T_3-T_4}=C\,\frac{\pi^2k_B^2T}{3h},
\end{align}
which is in direct agreement with the BB correspondence.
\section{Shot Noise at the QPC}\label{appB}
 Shot noise measurement involving counter-propagating edge modes of QH state at a QPC constriction yield the Fano factor, providing a measure of the quasi-particle charge involved in the tunneling at the QPC~\cite{Saminadayar1997,Picciotto1997,Reznikov1999,Chang2003,Dolev2008,Heiblum2020}. However, shot noise measurements in QH state at low temperature, supporting multiple edge modes, including upstream neutral modes, were reported to give a Fano factor equal to the bulk filling fraction~\cite{Bunched2003,bid2009}. Recent studies show that the presence of upstream neutral modes is crucial for obtaining a Fano factor equal to the bulk filling fraction~\cite{Gefen_2021,Biswas2022,Sabo2017,MannaNoise2024}. This process assumes the inter-edge charge equilibration together with the creation and subsequent decay of neutral excitations (neutralons) generated during equilibration. Within this theoretical framework, we compute the shot noise for the reconstructed edge structure $(+1/3, 0, +2/3)$. The resulting noise arises from the interplay between counter-propagating charge and neutral modes under the assumption of full charge equilibration. We analyze both on the plateau and weak back scattering regimes near the QPC conductance of $(1/3) e^2/h$.
 
\begin{figure}[t]
    \centering
    \includegraphics[width=0.98\linewidth]{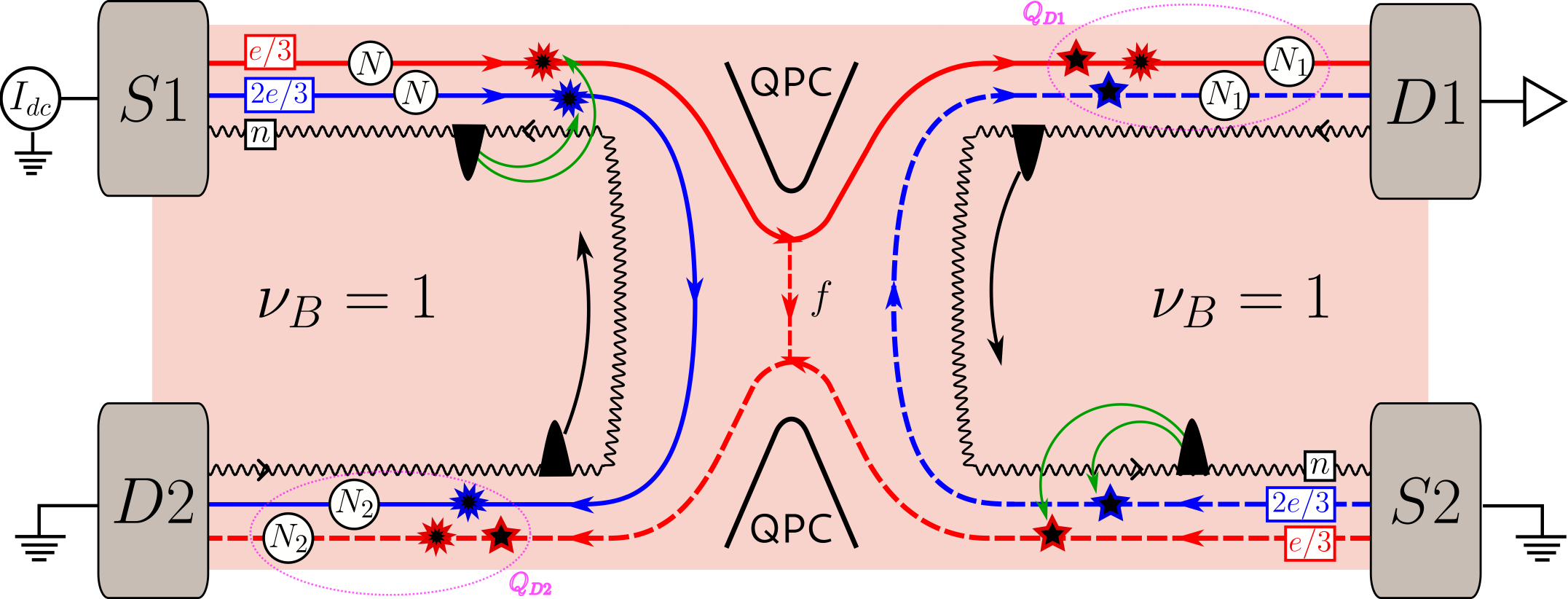}
    \caption{Schematic of a quantum Hall system at a bulk filling fraction of $1$, featuring the reconstructed edge structure $(+1/3, 0, +2/3)$. Charge modes from the biased source S1 (grounded source S2) are shown as solid (dashed) lines, with inner $2e/3$ (blue) and outer $e/3$ (red) quasi-particle charges. Upstream neutral modes are depicted by wavy lines. A fraction $f$ of the $e/3$ mode is reflected at the QPC toward drain D2 (D1) from S1 (S2). As the modes propagate toward the drains, inter-mode charge equilibration generates neutralons or anti-neutralons (black shaded areas). These neutral excitations propagate upstream across the QPC toward S2 and decay away from the QPC into randomized charge excitations (illustrated by red and blue star). Similarly, neutral excitations generated near D2 propagate upstream toward S1 and decay into charge excitations (red and blue multi-pointed stars). These decay processes generate shot noise at drains D1 and D2, even on the conductance plateau. The total charge reaching D1 and D2 is denoted by $Q_{D1}$ and $Q_{D2}$, respectively.}
    \label{fig:noise}
\end{figure}

In fig.~\ref{fig:noise}, we consider the source S1 emits $N$ quasi-particles into both charge mode in time $\tau$, resulting in a total injected current $I_{dc}=(1/3)eN/\tau+(2/3)eN/\tau=eN/\tau$. When the QPC is tuned to the $(1/3) e^2/h$ conductance plateau, the inner $2e/3$ charge mode is fully reflected, while the outer $e/3$ charge edge mode is fully transmitted. Away from the plateau, let $f$ denote the fraction of the $1/3$ mode that is reflected toward D2 (D1) when injected from S1 (S2). After transmission through the QPC, the `hot' charge modes injected from the voltage-biased source S1 (solid lines in Fig.~\ref{fig:noise}) and the `cold' charge modes injected from the grounded source S2 (dashed lines) propagate parallel to each other toward D1. During the time interval $\tau$, a charge $eN(1-f)/3$ is transmitted to D1 through the partially transmitting $1/3$ mode for $f\neq 0$. Near D1, the co-propagating $1/3$ and $2/3$ modes equilibrate via inter-mode tunneling processes that redistribute charge between them while conserving the total current entering D1. As a result, each mode carries $N_1=N(1-f)/3$ quasi-particles after equilibration. Similarly, a 2/3 and a partially reflected 1/3 mode carry a total charge $(2N/3+Nf/3)e$ charge to D2 and conservation of total current fixes the quasi-particle number in each mode after equilibration near D2 to be $N_2=N(2+f)/3$.

The equilibration near D1 (D2) produces neutralons (or anti-neutralons) in the inner upstream neutral mode that fully reflect through the QPC and propagate upstream toward S2 (S1). These neutralons subsequently decay (the upper-left and lower-right edge segments are assumed to be sufficiently long) into randomized quasi-particle and quasi-hole pairs that flow back to the QPC and partition into the drains. While this stochastic process leaves the average current invariant, it generates shot noise.

Let $a_i^{(\alpha)}$ and $b_i^{(\alpha)}$ denote the stochastic excitations near S2 and S1, respectively. These variables take values of $\pm 1$ corresponding to a quasi-particle/quasi-hole with equal probability and $\alpha$ indexes the modes (inner to outer), 
\begin{table}[h]
    \centering
   \begin{tabular}{|l|c|c|}
       \hline
        \textbf{Location} & \textbf{Inner Mode ($2/3$)} & \textbf{Outer Mode ($1/3$)} \\
        \hline
        Near S1 & $N - N_2$ & $N_2 - Nf$ \\
        \hline
        Near S2 & $N_1$     & $N(1-f) - N_1$ \\
        \hline
    \end{tabular}
        \caption{Stochastic excitations in the inner $2/3$ and outer $1/3$ modes prior to reaching the QPC.}   \label{tab:excitations}
\end{table}
while $i$ represents the chronological sequence of pulses. Table~III lists the resulting number of such excitations generated in each charge mode near the two sources. Among the excitations in the outer $1/3$ mode, a fraction $f$ originating near S2 (S1) reaches D1 (D2) due to stochastic tunneling events across the QPC, which are represented by the random variables $c_i$ ($d_i$). Thus the charge arriving at drain D1 and D2 is
\begin{align}
Q_{D1}=&N_1 e+ \frac{e}{3}\sum_{i=1}^{N_2-Nf} b_i^{(2)}-\frac{e}{3}\sum_{i=1}^{(N_2-Nf)f} d_i+\nonumber\\
&\quad\quad\quad\quad\quad\frac{2e}{3}\sum_{i=1}^{N_1} a_i^{(1)} +\frac{e}{3}\sum_{i=1}^{(N(1-f)-N_1)f}c_i,\\
Q_{D2}=&N_2 e+\frac{e}{3}\sum_{i=1}^{(N_2-Nf)f} d_i+\frac{2e}{3}\sum_{i=1}^{N-N_2} b_i^{(1)} +\nonumber\\
&\quad\quad\frac{e}{3}\sum_{i=1}^{N(1-f)-N_1}a_i^{(2)}-\frac{e}{3}\sum_{i=1}^{(N(1-f)-N_1)f}c_i.
\end{align}
Here, $\langle Q_{D1}\rangle=eN(1-f)/3$ and $\langle Q_{D2}\rangle=eN(2+f)/3$. With the following correlation properties
\begin{align}
    &\langle a_i^{(\alpha)} a_j^{(\beta)} \rangle=\langle b_i^{(\alpha)} b_j^{(\beta)} \rangle=\delta_{i,j}\delta_{\alpha,\beta}, \nonumber\\
    &\langle a_i^{(\alpha)} a_j^{(\beta)} \rangle=\langle b_i^{(\alpha)} b_j^{(\beta)} \rangle=-\delta_{i,j}\text{ for } \alpha\neq\beta,\nonumber\\
    & \langle a_i^{(\alpha)} c_j \rangle=\langle b_i^{(\alpha)} d_j \rangle=\delta_{i,j}\delta_{\alpha,2},\nonumber\\
    &\langle b_i^{(\alpha)} c_j \rangle=\langle a_i^{(\alpha)} d_j \rangle= \langle c_i\, d_j\rangle=0,
\end{align}
and writing $Q_X=\langle Q_X\rangle+\delta Q_X$ for each of these charges, the auto-correlation at D1 in the weak backscattering limit (small $f$) is
\begin{align}
    \langle (\delta Q_{D1})^2\rangle=&\frac{2Ne^2}{9} (1-f).
\end{align}
We next include the orthodox beam partitioning noise generated at the QPC. Consider a source injects $N$ quasi-particles of charge $e/3$ into a $1/3$ edge mode over a time interval $\tau$. Each quasi-particle is reflected across the QPC with probability $f$. To represent these mutually exclusive outcomes, introduce a binary variable $x_i$ corresponding to transmission with respect to the QPC as:
\begin{align}
x_i =&
\begin{cases}
1, & \text{if the quasi-particle is transmitted},\\
0, & \text{if it is reflected},
\end{cases}
\end{align}
Since a particle must be either transmitted or reflected, the corresponding variable for reflection is simply $y_i = 1 - x_i$. As binary variables, they satisfy $x_i^2 = x_i$ and $y_i^2 = y_i$. Their expectation values are: 
\begin{align}
\langle x_i\rangle&=1\times (1-f)+0\times f=1-f ,\nonumber\\ \langle y_i\rangle&=0\times (1-f)+1\times f=f.
\end{align}
The total transmitted and reflected charges are then given by:
\begin{equation}
Q_T = \frac{e}{3}\sum_{i=1}^N x_i, \qquad
Q_R = \frac{e}{3}\sum_{i=1}^N y_i .
\end{equation}
Defining the charge fluctuations as $\delta Q_{T,R}=Q_{T,R}-\langle Q_{T,R}\rangle$, and assuming statistically independent tunneling events, the variance of the transmitted charge is given by:
\begin{align}
\langle (\delta Q_T)^2\rangle
&=
\frac{e^2}{9}\sum_{i=1}^N \bigl(\langle x_i^2\rangle-\langle x_i\rangle^2\bigr) =\frac{Ne^2}{9}f(1-f),\label{eq:auto_ortho}
\end{align}
An identical result follows for the reflected charge,
\begin{align}
\langle(\delta Q_R)^2\rangle=\frac{Ne^2}{9}f(1-f).
\end{align}
The cross correlation between the transmitted and reflected charges is
\begin{equation}
\langle \delta Q_T\, \delta Q_R\rangle
=\frac{e^2}{9}
\bigg\langle
\sum_{i=1}^N \delta x_i
\sum_{j=1}^N \delta y_j
\bigg\rangle ,
\end{equation}
where $\delta x_i=x_i-\langle x_i\rangle$ and $\delta y_j = y_j - \langle y_j\rangle$. Because the tunneling events are statistically independent, the off-diagonal terms ($i \neq j$) vanish. With $\delta x_i=-\delta y_i$ for each quasi-particle, we obtain
\begin{equation}
\langle \delta Q_T \,\delta Q_R\rangle
=-\frac{e^2}{9}\sum_{i=1}^N
\langle (\delta x_i)^2\rangle
=-\frac{Ne^2}{9} f(1-f).\label{eq:cross_ortho}
\end{equation}
Combining the noise generated by the equilibration processes with the orthodox beam-partitioning contribution from eq.~(\ref{eq:auto_ortho})
\begin{align}
   \langle( \delta Q_{D1})^2\rangle=\frac{Ne^2}{9}(1-f)(2+f).
\end{align}
Since $I_{dc}=eN/\tau$ and the transmission parameter of the QPC is $t=(1-f)/3$, the auto-correlation Fano factor at D1 is defined as:
\begin{align}
    F_1=\frac{ \langle( \delta Q_{D1})^2\rangle}{e\,\tau I_{dc} t(1-t)}=1,
\end{align}
Remarkably, this result is equal to the bulk filling fraction both on the conductance plateau and in weak back scattering limit. An identical analysis for the auto-correlation at D2 similarly yields $F_2 = 1$. For the cross correlation calculation
\begin{align}
    \langle \delta Q_{D1}\, \delta Q_{D2}\rangle=-\frac{4e^2N}{27}(1-f).
\end{align}
Adding this to the orthodox beam-partitioning contribution of $-f(1-f)Ne^2/9$ from eq.~(\ref{eq:cross_ortho}), we obtain the total cross-correlation:
\begin{equation}
    \langle \delta Q_{D1}\, \delta Q_{D2}\rangle = -\frac{Ne^2}{27}(1-f)(4+3f).
\end{equation}
Consequently, dividing by the same factor used for the auto-correlation, the Fano factor for the cross-correlation between D1 and D2 results
\begin{align}
    F_c=-\frac{1}{3}\left(\frac{4+3f}{2+f}\right).
\end{align}
In the specific case where the QPC is tuned to the conductance plateau ($f=0$), this reduces to $F_c = -2/3$. Furthermore, global charge conservation dictates that the total charge injected from S1 and S2 (denoted by $Q_{S1}$ and $Q_{S2}$, respectively) must equal the total charge collected at D1 and D2: $Q_{S1}+Q_{S2}=Q_{D1}+Q_{D2}$. The injected charges are given by:
\begin{align}
    Q_{S1}&=Ne+\frac{e}{3} \sum_{i=1}^{N_2-Nf}b_i^{(2)}+\frac{2e}{3} \sum_{i=1}^{N-N_2}b_i^{(1)}\\
     Q_{S2}&=\frac{e}{3} \sum_{i=1}^{N(1-f)-N_1}a_i^{(2)}+\frac{2e}{3} \sum_{i=1}^{N_1}a_i^{(1)}.
\end{align}
Given the expectation values $\langle Q_{S1}\rangle= Ne$, $\langle Q_{S2}\rangle=0$, $\langle Q_{D1}\rangle= Ne(1-f)/3$, and $\langle Q_{D2}\rangle= Ne(2+f)/3$,the fluctuating components must satisfy
\begin{align}
    \delta Q_{S1}+ \delta Q_{S2}-(\delta Q_{D1}+\delta Q_{D2})=0.
\end{align}
Squaring and statistically averaging both sides give all auto and cross correlation noise terms (source–source, drain–drain, and source–drain) as follows
\begin{align}
  \langle ( \delta& Q_{S1})^2\rangle+ \langle ( \delta Q_{S2})^2\rangle+ \langle ( \delta Q_{D1})^2\rangle+\langle ( \delta Q_{D2})^2\rangle+\nonumber\\
&2\langle  \delta Q_{S1}\, \delta Q_{S2}\rangle-2\langle  \delta Q_{S1}\, \delta Q_{D1}\rangle-2 \langle \delta Q_{S1}\, \delta Q_{D2}\rangle-\nonumber\\
  &2 \langle \delta Q_{S2}\, \delta Q_{D1}\rangle-2 \langle \delta Q_{S2}\, \delta Q_{D2}\rangle+2 \langle \delta Q_{D1}\, \delta Q_{D2}\rangle=0.\label{eq:all_noise}
\end{align}
Each of these correlation terms yields
\begin{align}
   & \langle ( \delta Q_{S1})^2\rangle=\frac{2e^2N}{27}(1-f),\quad\langle ( \delta Q_{S2})^2\rangle=\frac{2e^2N}{27}(1-f),\nonumber\\
   &\langle ( \delta Q_{D1})^2\rangle=\frac{2e^2N}{9}(1-f),\quad \langle ( \delta Q_{D2})^2\rangle=\frac{2e^2N}{9}(1-f),\nonumber\\
   & \langle  \delta Q_{S1}\, \delta Q_{S2}\rangle=0,\quad\quad   \langle  \delta Q_{S1}\, \delta Q_{D1}\rangle=-\frac{2e^2N}{27}f(1-f),\nonumber\\
    & \langle  \delta Q_{S1}\, \delta Q_{D2}\rangle=\frac{2e^2N}{27}(1-f^2),\nonumber\\
    &\langle  \delta Q_{S2}\, \delta Q_{D1}\rangle=\frac{2e^2N}{27}(1-f^2), \nonumber\\
    & \langle  \delta Q_{S2}\, \delta Q_{D2}\rangle=-\frac{2e^2N}{27}f(1-f),\nonumber\\
    &\langle  \delta Q_{D1}\, \delta Q_{D2}\rangle=-\frac{4e^2N}{27}(1-f).\nonumber
\end{align}
Summing these explicitly calculated terms confirms that the LHS of eq.~(\ref{eq:all_noise}) exactly vanishes. This provides a rigorous consistency check in accordance with charge conservation.
\bibliography{citations}

@article{wen1991,
  title = {Gapless boundary excitations in the quantum Hall states and in the chiral spin states},
  author = {Wen, X. G.},
  journal = {Phys. Rev. B},
  volume = {43},
  issue = {13},
  pages = {11025--11036},
  numpages = {0},
  year = {1991},
  month = {May},
  publisher = {American Physical Society},
  doi = {10.1103/PhysRevB.43.11025},
  url = {https://link.aps.org/doi/10.1103/PhysRevB.43.11025}
}

@Article{Bid2010,
author={Bid, Aveek
and Ofek, N.
and Inoue, H.
and Heiblum, M.
and Kane, C. L.
and Umansky, V.
and Mahalu, D.},
title={Observation of neutral modes in the fractional quantum Hall regime},
journal={Nature},
year={2010},
month={Jul},
day={01},
volume={466},
number={7306},
pages={585-590},
issn={1476-4687},
doi={10.1038/nature09277},
url={https://doi.org/10.1038/nature09277}
}

@article{Deviatov,
  title = {Energy Transport by Neutral Collective Excitations at the Quantum Hall Edge},
  author = {Deviatov, E. V. and Lorke, A. and Biasiol, G. and Sorba, L.},
  journal = {Phys. Rev. Lett.},
  volume = {106},
  issue = {25},
  pages = {256802},
  numpages = {4},
  year = {2011},
  month = {Jun},
  publisher = {American Physical Society},
  doi = {10.1103/PhysRevLett.106.256802},
  url = {https://link.aps.org/doi/10.1103/PhysRevLett.106.256802}
}

@article{Yaron2012,
  title = {Upstream Neutral Modes in the Fractional Quantum Hall Effect Regime: Heat Waves or Coherent Dipoles},
  author = {Gross, Yaron and Dolev, Merav and Heiblum, Moty and Umansky, Vladimir and Mahalu, Diana},
  journal = {Phys. Rev. Lett.},
  volume = {108},
  issue = {22},
  pages = {226801},
  numpages = {5},
  year = {2012},
  month = {May},
  publisher = {American Physical Society},
  doi = {10.1103/PhysRevLett.108.226801},
  url = {https://link.aps.org/doi/10.1103/PhysRevLett.108.226801}
}

@Article{Gurman2012,
author={Gurman, I.
and Sabo, R.
and Heiblum, M.
and Umansky, V.
and Mahalu, D.},
title={Extracting net current from an upstream neutral mode in the fractional quantum Hall regime},
journal={Nature Communications},
year={2012},
month={Dec},
day={18},
volume={3},
number={1},
pages={1289},
issn={2041-1723},
doi={10.1038/ncomms2305},
url={https://doi.org/10.1038/ncomms2305}
}

@Article{Venkatachalam2012,
author={Venkatachalam, Vivek
and Hart, Sean
and Pfeiffer, Loren
and West, Ken
and Yacoby, Amir},
title={Local thermometry of neutral modes on the quantum Hall edge},
journal={Nature Physics},
year={2012},
month={Sep},
day={01},
volume={8},
number={9},
pages={676-681},
issn={1745-2481},
doi={10.1038/nphys2384},
url={https://doi.org/10.1038/nphys2384}
}

@Article{Inoue2014,
author={Inoue, Hiroyuki
and Grivnin, Anna
and Ronen, Yuval
and Heiblum, Moty
and Umansky, Vladimir
and Mahalu, Diana},
title={Proliferation of neutral modes in fractional quantum Hall states},
journal={Nature Communications},
year={2014},
month={Jun},
day={06},
volume={5},
number={1},
pages={4067},
issn={2041-1723},
doi={10.1038/ncomms5067},
url={https://doi.org/10.1038/ncomms5067}
}

@article{Etyan_Sourin_2009,
  title = {Probing the Neutral Edge Modes in Transport across a Point Contact via Thermal Effects in the Read-Rezayi Non-Abelian Quantum Hall States},
  author = {Grosfeld, Eytan and Das, Sourin},
  journal = {Phys. Rev. Lett.},
  volume = {102},
  issue = {10},
  pages = {106403},
  numpages = {4},
  year = {2009},
  month = {Mar},
  publisher = {American Physical Society},
  doi = {10.1103/PhysRevLett.102.106403},
  url = {https://link.aps.org/doi/10.1103/PhysRevLett.102.106403}
}

@article{Plasmon_scattering2010,
  title = {Plasmon scattering approach to energy exchange and high-frequency noise in $\ensuremath{\nu}=2$ quantum Hall edge channels},
  author = {Degiovanni, P. and Grenier, Ch. and F\`eve, G. and Altimiras, C. and le Sueur, H. and Pierre, F.},
  journal = {Phys. Rev. B},
  volume = {81},
  issue = {12},
  pages = {121302},
  numpages = {4},
  year = {2010},
  month = {Mar},
  publisher = {American Physical Society},
  doi = {10.1103/PhysRevB.81.121302},
  url = {https://link.aps.org/doi/10.1103/PhysRevB.81.121302}
}

@article{Giovani_Sourin_2012,
  title = {Thermoelectric Probe for Neutral Edge Modes in the Fractional Quantum Hall Regime},
  author = {Viola, Giovanni and Das, Sourin and Grosfeld, Eytan and Stern, Ady},
  journal = {Phys. Rev. Lett.},
  volume = {109},
  issue = {14},
  pages = {146801},
  numpages = {5},
  year = {2012},
  month = {Oct},
  publisher = {American Physical Society},
  doi = {10.1103/PhysRevLett.109.146801},
  url = {https://link.aps.org/doi/10.1103/PhysRevLett.109.146801}
}

@article{Altimiras2012,
  title = {Chargeless Heat Transport in the Fractional Quantum Hall Regime},
  author = {Altimiras, C. and le Sueur, H. and Gennser, U. and Anthore, A. and Cavanna, A. and Mailly, D. and Pierre, F.},
  journal = {Phys. Rev. Lett.},
  volume = {109},
  issue = {2},
  pages = {026803},
  numpages = {5},
  year = {2012},
  month = {Jul},
  publisher = {American Physical Society},
  doi = {10.1103/PhysRevLett.109.026803},
  url = {https://link.aps.org/doi/10.1103/PhysRevLett.109.026803}
}

@Article{Banerjee2017,
author={Banerjee, Mitali
and Heiblum, Moty
and Rosenblatt, Amir
and Oreg, Yuval
and Feldman, Dima E.
and Stern, Ady
and Umansky, Vladimir},
title={Observed quantization of anyonic heat flow},
journal={Nature},
year={2017},
month={May},
day={01},
volume={545},
number={7652},
pages={75-79},
issn={1476-4687},
doi={10.1038/nature22052},
url={https://doi.org/10.1038/nature22052}
}

@Article{Banerjee2018,
author={Banerjee, Mitali
and Heiblum, Moty
and Umansky, Vladimir
and Feldman, Dima E.
and Oreg, Yuval
and Stern, Ady},
title={Observation of half-integer thermal Hall conductance},
journal={Nature},
year={2018},
month={Jul},
day={01},
volume={559},
number={7713},
pages={205-210},
issn={1476-4687},
doi={10.1038/s41586-018-0184-1},
url={https://doi.org/10.1038/s41586-018-0184-1}
}

@article{Anindya2021,
  title = {Vanishing Thermal Equilibration for Hole-Conjugate Fractional Quantum Hall States in Graphene},
  author = {Srivastav, Saurabh Kumar and Kumar, Ravi and Sp\aa{}nsl\"att, Christian and Watanabe, K. and Taniguchi, T. and Mirlin, Alexander D. and Gefen, Yuval and Das, Anindya},
  journal = {Phys. Rev. Lett.},
  volume = {126},
  issue = {21},
  pages = {216803},
  numpages = {6},
  year = {2021},
  month = {May},
  publisher = {American Physical Society},
  doi = {10.1103/PhysRevLett.126.216803},
  url = {https://link.aps.org/doi/10.1103/PhysRevLett.126.216803}
}

@article{Anindya2019,
author = {Saurabh Kumar Srivastav  and Manas Ranjan Sahu  and K. Watanabe  and T. Taniguchi  and Sumilan Banerjee  and Anindya Das },
title = {Universal quantized thermal conductance in graphene},
journal = {Science Advances},
volume = {5},
number = {7},
pages = {eaaw5798},
year = {2019},
doi = {10.1126/sciadv.aaw5798}}

@Article{Srivastav2022,
author={Srivastav, Saurabh Kumar
and Kumar, Ravi
and Sp{\aa}nsl{\"a}tt, Christian
and Watanabe, K.
and Taniguchi, T.
and Mirlin, Alexander D.
and Gefen, Yuval
and Das, Anindya},
title={Determination of topological edge quantum numbers of fractional quantum Hall phases by thermal conductance measurements},
journal={Nature Communications},
year={2022},
month={Sep},
day={03},
volume={13},
number={1},
pages={5185},
issn={2041-1723},
doi={10.1038/s41467-022-32956-z},
url={https://doi.org/10.1038/s41467-022-32956-z}
}

@article{temp_Enhancement,
  title = {Temperature Enhancement of Thermal Hall Conductance Quantization},
  author = {Fulga, I. C. and Oreg, Yuval and Mirlin, Alexander D. and Stern, Ady and Mross, David F.},
  journal = {Phys. Rev. Lett.},
  volume = {125},
  issue = {23},
  pages = {236802},
  numpages = {6},
  year = {2020},
  month = {Dec},
  publisher = {American Physical Society},
  doi = {10.1103/PhysRevLett.125.236802},
  url = {https://link.aps.org/doi/10.1103/PhysRevLett.125.236802}
}

@Article{Melcer2022,
author={Melcer, Ron Aharon
and Dutta, Bivas
and Sp{\aa}nsl{\"a}tt, Christian
and Park, Jinhong
and Mirlin, Alexander D.
and Umansky, Vladimir},
title={Absent thermal equilibration on fractional quantum Hall edges over macroscopic scale},
journal={Nature Communications},
year={2022},
month={Jan},
day={19},
volume={13},
number={1},
pages={376},
issn={2041-1723},
doi={10.1038/s41467-022-28009-0},
url={https://doi.org/10.1038/s41467-022-28009-0}
}

@Article{melcer2023direct,
author={Melcer, Ron Aharon
and Konyzheva, Sofia
and Heiblum, Moty
and Umansky, Vladimir},
title={Direct determination of the topological thermal conductance via local power measurement},
journal={Nature Physics},
year={2023},
month={Mar},
day={01},
volume={19},
number={3},
pages={327-332},
issn={1745-2481},
doi={10.1038/s41567-022-01885-5},
url={https://doi.org/10.1038/s41567-022-01885-5}
}

@Article{Melcer2024,
author={Melcer, Ron Aharon
and Gil, Avigail
and Paul, Arup Kumar
and Tiwari, Priya
and Umansky, Vladimir
and Heiblum, Moty
and Oreg, Yuval
and Stern, Ady
and Berg, Erez},
title={Heat conductance of the quantum Hall bulk},
journal={Nature},
year={2024},
month={Jan},
day={01},
volume={625},
number={7995},
pages={489-493},
issn={1476-4687},
doi={10.1038/s41586-023-06858-z},
url={https://doi.org/10.1038/s41586-023-06858-z}
}

@article{Chklovskii1992,
  title = {Electrostatics of edge channels},
  author = {Chklovskii, D. B. and Shklovskii, B. I. and Glazman, L. I.},
  journal = {Phys. Rev. B},
  volume = {46},
  issue = {7},
  pages = {4026--4034},
  numpages = {0},
  year = {1992},
  month = {Aug},
  publisher = {American Physical Society},
  doi = {10.1103/PhysRevB.46.4026},
  url = {https://link.aps.org/doi/10.1103/PhysRevB.46.4026}
}

@article{Halperin1993,
  title = {Electron-electron interactions and spontaneous spin polarization in quantum Hall edge states},
  author = {Dempsey, Jed and Gelfand, B. Y. and Halperin, B. I.},
  journal = {Phys. Rev. Lett.},
  volume = {70},
  issue = {23},
  pages = {3639--3642},
  numpages = {0},
  year = {1993},
  month = {Jun},
  publisher = {American Physical Society},
  doi = {10.1103/PhysRevLett.70.3639},
  url = {https://link.aps.org/doi/10.1103/PhysRevLett.70.3639}
}

@article{Wen_1994,
  title = {Sharp and smooth boundaries of quantum Hall liquids},
  author = {Chamon, C. de C. and Wen, X. G.},
  journal = {Phys. Rev. B},
  volume = {49},
  issue = {12},
  pages = {8227--8241},
  numpages = {0},
  year = {1994},
  month = {Mar},
  publisher = {American Physical Society},
  doi = {10.1103/PhysRevB.49.8227},
  url = {https://link.aps.org/doi/10.1103/PhysRevB.49.8227}
}

@article{Meir_1994,
  title = {Composite edge states in the \ensuremath{\nu}=2/3 fractional quantum Hall regime},
  author = {Meir, Yigal},
  journal = {Phys. Rev. Lett.},
  volume = {72},
  issue = {16},
  pages = {2624--2627},
  numpages = {0},
  year = {1994},
  month = {Apr},
  publisher = {American Physical Society},
  doi = {10.1103/PhysRevLett.72.2624},
  url = {https://link.aps.org/doi/10.1103/PhysRevLett.72.2624}
}

@article{Chklovskii1995,
  title = {Structure of fractional edge states: A composite-fermion approach},
  author = {Chklovskii, Dmitri B.},
  journal = {Phys. Rev. B},
  volume = {51},
  issue = {15},
  pages = {9895--9902},
  numpages = {0},
  year = {1995},
  month = {Apr},
  publisher = {American Physical Society},
  doi = {10.1103/PhysRevB.51.9895},
  url = {https://link.aps.org/doi/10.1103/PhysRevB.51.9895}
}

@article{Kun2002,
  title = {Reconstruction of Fractional Quantum Hall Edges},
  author = {Wan, Xin and Yang, Kun and Rezayi, E. H.},
  journal = {Phys. Rev. Lett.},
  volume = {88},
  issue = {5},
  pages = {056802},
  numpages = {4},
  year = {2002},
  month = {Jan},
  publisher = {American Physical Society},
  doi = {10.1103/PhysRevLett.88.056802},
  url = {https://link.aps.org/doi/10.1103/PhysRevLett.88.056802}
}

@article{Kun2003,
  title = {Edge reconstruction in the fractional quantum Hall regime},
  author = {Wan, Xin and Rezayi, E. H. and Yang, Kun},
  journal = {Phys. Rev. B},
  volume = {68},
  issue = {12},
  pages = {125307},
  numpages = {12},
  year = {2003},
  month = {Sep},
  publisher = {American Physical Society},
  doi = {10.1103/PhysRevB.68.125307},
  url = {https://link.aps.org/doi/10.1103/PhysRevB.68.125307}
}

@article{Ganpathy2003,
  title = {Edge reconstructions in fractional quantum Hall systems},
  author = {Joglekar, Yogesh N. and Nguyen, Hoang K. and Murthy, Ganpathy},
  journal = {Phys. Rev. B},
  volume = {68},
  issue = {3},
  pages = {035332},
  numpages = {9},
  year = {2003},
  month = {Jul},
  publisher = {American Physical Society},
  doi = {10.1103/PhysRevB.68.035332},
  url = {https://link.aps.org/doi/10.1103/PhysRevB.68.035332}
}

@article{Meir2013,
  title = {Edge Reconstruction in the $\ensuremath{\nu}\mathbf{=}2/3$ Fractional Quantum Hall State},
  author = {Wang, Jianhui and Meir, Yigal and Gefen, Yuval},
  journal = {Phys. Rev. Lett.},
  volume = {111},
  issue = {24},
  pages = {246803},
  numpages = {5},
  year = {2013},
  month = {Dec},
  publisher = {American Physical Society},
  doi = {10.1103/PhysRevLett.111.246803},
  url = {https://link.aps.org/doi/10.1103/PhysRevLett.111.246803}
}

@article{KFP_1994,
  title = {Randomness at the edge: Theory of quantum Hall transport at filling \ensuremath{\nu}=2/3},
  author = {Kane, C. L. and Fisher, Matthew P. A. and Polchinski, J.},
  journal = {Phys. Rev. Lett.},
  volume = {72},
  issue = {26},
  pages = {4129--4132},
  numpages = {0},
  year = {1994},
  month = {Jun},
  publisher = {American Physical Society},
  doi = {10.1103/PhysRevLett.72.4129},
  url = {https://link.aps.org/doi/10.1103/PhysRevLett.72.4129}
}

@article{kane1995,
  title = {Impurity scattering and transport of fractional quantum Hall edge states},
  author = {Kane, C. L. and Fisher, Matthew P. A.},
  journal = {Phys. Rev. B},
  volume = {51},
  issue = {19},
  pages = {13449--13466},
  numpages = {0},
  year = {1995},
  month = {May},
  publisher = {American Physical Society},
  doi = {10.1103/PhysRevB.51.13449},
  url = {https://link.aps.org/doi/10.1103/PhysRevB.51.13449}
}

@article{PROTOPOPOV2017287,
title = {Transport in a disordered $\ensuremath{\nu}=2/3$ fractional quantum Hall junction},
journal = {Annals of Physics},
volume = {385},
pages = {287-327},
year = {2017},
issn = {0003-4916},
doi = {https://doi.org/10.1016/j.aop.2017.07.015},
url = {https://www.sciencedirect.com/science/article/pii/S0003491617302142},
author = {I.V. Protopopov and Yuval Gefen and A.D. Mirlin}
}

@article{Park2019,
  title = {Noise on complex quantum Hall edges: Chiral anomaly and heat diffusion},
  author = {Park, Jinhong and Mirlin, Alexander D. and Rosenow, Bernd and Gefen, Yuval},
  journal = {Phys. Rev. B},
  volume = {99},
  issue = {16},
  pages = {161302},
  numpages = {6},
  year = {2019},
  month = {Apr},
  publisher = {American Physical Society},
  doi = {10.1103/PhysRevB.99.161302},
  url = {https://link.aps.org/doi/10.1103/PhysRevB.99.161302}
}

@article{mirlin2019,
  title = {Topological Classification of Shot Noise on Fractional Quantum Hall Edges},
  author = {Sp\aa{}nsl\"att, Christian and Park, Jinhong and Gefen, Yuval and Mirlin, Alexander D.},
  journal = {Phys. Rev. Lett.},
  volume = {123},
  issue = {13},
  pages = {137701},
  numpages = {6},
  year = {2019},
  month = {Sep},
  publisher = {American Physical Society},
  doi = {10.1103/PhysRevLett.123.137701},
  url = {https://link.aps.org/doi/10.1103/PhysRevLett.123.137701}
}

@article{Park2020,
  title = {Noise on the non-Abelian $\ensuremath{\nu}=5/2$ Fractional Quantum Hall Edge},
  author = {Park, Jinhong and Sp\aa{}nsl\"att, Christian and Gefen, Yuval and Mirlin, Alexander D.},
  journal = {Phys. Rev. Lett.},
  volume = {125},
  issue = {15},
  pages = {157702},
  numpages = {7},
  year = {2020},
  month = {Oct},
  publisher = {American Physical Society},
  doi = {10.1103/PhysRevLett.125.157702},
  url = {https://link.aps.org/doi/10.1103/PhysRevLett.125.157702}
}

@article{Yuval_Conductance_plateau,
  title = {Conductance plateaus and shot noise in fractional quantum Hall point contacts},
  author = {Sp\aa{}nsl\"att, Christian and Park, Jinhong and Gefen, Yuval and Mirlin, Alexander D.},
  journal = {Phys. Rev. B},
  volume = {101},
  issue = {7},
  pages = {075308},
  numpages = {20},
  year = {2020},
  month = {Feb},
  publisher = {American Physical Society},
  doi = {10.1103/PhysRevB.101.075308},
  url = {https://link.aps.org/doi/10.1103/PhysRevB.101.075308}
}

@article{yuval_contacts,
  title = {Contacts, equilibration, and interactions in fractional quantum Hall edge transport},
  author = {Sp\aa{}nsl\"att, C. and Gefen, Yuval and Gornyi, I. V. and Polyakov, D. G.},
  journal = {Phys. Rev. B},
  volume = {104},
  issue = {11},
  pages = {115416},
  numpages = {28},
  year = {2021},
  month = {Sep},
  publisher = {American Physical Society},
  doi = {10.1103/PhysRevB.104.115416},
  url = {https://link.aps.org/doi/10.1103/PhysRevB.104.115416}
}

@article{Gefen_2022,
  title = {Emergence of Neutral Modes in Laughlin-like Fractional Quantum Hall Phases},
  author = {Khanna, Udit and Goldstein, Moshe and Gefen, Yuval},
  journal = {Phys. Rev. Lett.},
  volume = {129},
  issue = {14},
  pages = {146801},
  numpages = {7},
  year = {2022},
  month = {Sep},
  publisher = {American Physical Society},
  doi = {10.1103/PhysRevLett.129.146801},
  url = {https://link.aps.org/doi/10.1103/PhysRevLett.129.146801}
}

@article{Maiti_2020,
  title = {Magnetic-Field-Dependent Equilibration of Fractional Quantum Hall Edge Modes},
  author = {Maiti, Tanmay and Agarwal, Pooja and Purkait, Suvankar and Sreejith, G. J. and Das, Sourin and Biasiol, Giorgio and Sorba, Lucia and Karmakar, Biswajit},
  journal = {Phys. Rev. Lett.},
  volume = {125},
  issue = {7},
  pages = {076802},
  numpages = {6},
  year = {2020},
  month = {Aug},
  publisher = {American Physical Society},
  doi = {10.1103/PhysRevLett.125.076802},
  url = {https://link.aps.org/doi/10.1103/PhysRevLett.125.076802}
}

@article{manna2024,
  title = {Multiple Mechanisms for Emerging Conductance Plateaus in Fractional Quantum Hall States},
  author = {Manna, Sourav and Das, Ankur and Gefen, Yuval and Goldstein, Moshe},
  journal = {Phys. Rev. Lett.},
  volume = {134},
  issue = {25},
  pages = {256503},
  numpages = {7},
  year = {2025},
  month = {Jun},
  publisher = {American Physical Society},
  doi = {10.1103/p4y5-trph},
  url = {https://link.aps.org/doi/10.1103/p4y5-trph}
}

@article{manna2023experimentallymotivatedorderlength,
  title = {Experimentally motivated order of length scales affect shot noise},
  author = {Manna, Sourav and Das, Ankur},
  journal = {Phys. Rev. B},
  volume = {112},
  issue = {19},
  pages = {195128},
  numpages = {11},
  year = {2025},
  month = {Nov},
  publisher = {American Physical Society},
  doi = {10.1103/lzph-3yzb},
  url = {https://link.aps.org/doi/10.1103/lzph-3yzb}
}

@article{Ankur_5by2,
  title = {Full Classification of Transport on an Equilibrated $5/2$ Edge via Shot Noise},
  author = {Manna, Sourav and Das, Ankur and Goldstein, Moshe and Gefen, Yuval},
  journal = {Phys. Rev. Lett.},
  volume = {132},
  issue = {13},
  pages = {136502},
  numpages = {7},
  year = {2024},
  month = {Mar},
  publisher = {American Physical Society},
  doi = {10.1103/PhysRevLett.132.136502},
  url = {https://link.aps.org/doi/10.1103/PhysRevLett.132.136502}
}

@misc{pandey2024halfquantizedhallplateausconfined,
      title={Half-quantized Hall Plateaus in the Confined Geometry of Graphene}, 
      author={Preeti Pandey and Sourav Manna and Kristiana N. Frei and Jerin Saji and Anne Denis and Alexander Savin and Kenji Watanabe and Takashi Taniguchi and Pertti J. Hakonen and Ankur Das and Manohar Kumar},
      year={2024},
      eprint={2410.03896},
      archivePrefix={arXiv},
      primaryClass={cond-mat.mes-hall},
      url={https://arxiv.org/abs/2410.03896}, 
}

@article{Tsui_5by2,
  title = {Observation of an even-denominator quantum number in the fractional quantum Hall effect},
  author = {Willett, R. and Eisenstein, J. P. and St\"ormer, H. L. and Tsui, D. C. and Gossard, A. C. and English, J. H.},
  journal = {Phys. Rev. Lett.},
  volume = {59},
  issue = {15},
  pages = {1776--1779},
  numpages = {0},
  year = {1987},
  month = {Oct},
  publisher = {American Physical Society},
  doi = {10.1103/PhysRevLett.59.1776},
  url = {https://link.aps.org/doi/10.1103/PhysRevLett.59.1776}
}

@article{Yutushi,
  title = {Identifying the $\ensuremath{\nu}=\frac{5}{2}$ Topological Order through Charge Transport Measurements},
  author = {Yutushui, Misha and Stern, Ady and Mross, David F.},
  journal = {Phys. Rev. Lett.},
  volume = {128},
  issue = {1},
  pages = {016401},
  numpages = {6},
  year = {2022},
  month = {Jan},
  publisher = {American Physical Society},
  doi = {10.1103/PhysRevLett.128.016401},
  url = {https://link.aps.org/doi/10.1103/PhysRevLett.128.016401}
}

@article{Park2024,
  title = {Fingerprints of Anti-Pfaffian Topological Order in Quantum Point Contact Transport},
  author = {Park, Jinhong and Sp\aa{}nsl\"att, Christian and Mirlin, Alexander D.},
  journal = {Phys. Rev. Lett.},
  volume = {132},
  issue = {25},
  pages = {256601},
  numpages = {7},
  year = {2024},
  month = {Jun},
  publisher = {American Physical Society},
  doi = {10.1103/PhysRevLett.132.256601},
  url = {https://link.aps.org/doi/10.1103/PhysRevLett.132.256601}
}

@article{Moty_2019,
  title = {Melting of Interference in the Fractional Quantum Hall Effect: Appearance of Neutral Modes},
  author = {Bhattacharyya, Rajarshi and Banerjee, Mitali and Heiblum, Moty and Mahalu, Diana and Umansky, Vladimir},
  journal = {Phys. Rev. Lett.},
  volume = {122},
  issue = {24},
  pages = {246801},
  numpages = {5},
  year = {2019},
  month = {Jun},
  publisher = {American Physical Society},
  doi = {10.1103/PhysRevLett.122.246801},
  url = {https://link.aps.org/doi/10.1103/PhysRevLett.122.246801}
}

@article{Gefen_2016,
  title = {Suppression of Interference in Quantum Hall Mach-Zehnder Geometry by Upstream Neutral Modes},
  author = {Goldstein, Moshe and Gefen, Yuval},
  journal = {Phys. Rev. Lett.},
  volume = {117},
  issue = {27},
  pages = {276804},
  numpages = {6},
  year = {2016},
  month = {Dec},
  publisher = {American Physical Society},
  doi = {10.1103/PhysRevLett.117.276804},
  url = {https://link.aps.org/doi/10.1103/PhysRevLett.117.276804}
}

@article{Paradiso_2012,
  title = {Imaging Fractional Incompressible Stripes in Integer Quantum Hall Systems},
  author = {Paradiso, Nicola and Heun, Stefan and Roddaro, Stefano and Sorba, Lucia and Beltram, Fabio and Biasiol, Giorgio and Pfeiffer, L. N. and West, K. W.},
  journal = {Phys. Rev. Lett.},
  volume = {108},
  issue = {24},
  pages = {246801},
  numpages = {5},
  year = {2012},
  month = {Jun},
  publisher = {American Physical Society},
  doi = {10.1103/PhysRevLett.108.246801},
  url = {https://link.aps.org/doi/10.1103/PhysRevLett.108.246801}
}

@article{Pascher_2014,
  title = {Imaging the Conductance of Integer and Fractional Quantum Hall Edge States},
  author = {Pascher, Nikola and R\"ossler, Clemens and Ihn, Thomas and Ensslin, Klaus and Reichl, Christian and Wegscheider, Werner},
  journal = {Phys. Rev. X},
  volume = {4},
  issue = {1},
  pages = {011014},
  numpages = {8},
  year = {2014},
  month = {Jan},
  publisher = {American Physical Society},
  doi = {10.1103/PhysRevX.4.011014},
  url = {https://link.aps.org/doi/10.1103/PhysRevX.4.011014}
}

@article{shtanko2014nonequilibrium,
  title = {Nonequilibrium noise in transport across a tunneling contact between $\ensuremath{\nu}=\frac{2}{3}$ fractional quantum Hall edges},
  author = {Shtanko, O. and Snizhko, K. and Cheianov, V.},
  journal = {Phys. Rev. B},
  volume = {89},
  issue = {12},
  pages = {125104},
  numpages = {12},
  year = {2014},
  month = {Mar},
  publisher = {American Physical Society},
  doi = {10.1103/PhysRevB.89.125104},
  url = {https://link.aps.org/doi/10.1103/PhysRevB.89.125104}
}

@article{AHM1987,
  title = {Edge states, transmission matrices, and the Hall resistance},
  author = {Streda, P. and Kucera, J. and MacDonald, A. H.},
  journal = {Phys. Rev. Lett.},
  volume = {59},
  issue = {17},
  pages = {1973--1975},
  numpages = {0},
  year = {1987},
  month = {Oct},
  publisher = {American Physical Society},
  doi = {10.1103/PhysRevLett.59.1973},
  url = {https://link.aps.org/doi/10.1103/PhysRevLett.59.1973}
}

@Article{roy2025,
author={Roy, Ujjal
and Manna, Sourav
and Chakraborty, Souvik
and Watanabe, Kenji
and Taniguchi, Takashi
and Das, Ankur
and Goldstein, Moshe
and Gefen, Yuval
and Das, Anindya},
title={Half-integer thermal conductance in integer quantum Hall states},
journal={Nature Communications},
year={2026},
month={Feb},
day={17},
volume={17},
number={1},
pages={2853},
issn={2041-1723},
doi={10.1038/s41467-026-69659-8},
url={https://doi.org/10.1038/s41467-026-69659-8}
}

@Article{Bocquillon2013,
author={Bocquillon, E.
and Freulon, V.
and Berroir, J.-.. M.
and Degiovanni, P.
and Pla{\c{c}}ais, B.
and Cavanna, A.
and Jin, Y.
and F{\`e}ve, G.},
title={Separation of neutral and charge modes in one-dimensional chiral edge channels},
journal={Nature Communications},
year={2013},
month={May},
day={14},
volume={4},
number={1},
pages={1839},
issn={2041-1723},
doi={10.1038/ncomms2788},
url={https://doi.org/10.1038/ncomms2788}
}

@article{Neutral_vel2,
  title = {Long tunneling contact as a probe of fractional quantum Hall neutral edge modes},
  author = {Overbosch, B. J. and Chamon, Claudio},
  journal = {Phys. Rev. B},
  volume = {80},
  issue = {3},
  pages = {035319},
  numpages = {5},
  year = {2009},
  month = {Jul},
  publisher = {American Physical Society},
  doi = {10.1103/PhysRevB.80.035319},
  url = {https://link.aps.org/doi/10.1103/PhysRevB.80.035319}
}

@article{Feldman,
  title = {Different Fractional Charges from Auto- and Cross-Correlation Noise in Quantum Hall States without Upstream Modes},
  author = {Batra, Navketan and Feldman, D. E.},
  journal = {Phys. Rev. Lett.},
  volume = {132},
  issue = {22},
  pages = {226601},
  numpages = {6},
  year = {2024},
  month = {May},
  publisher = {American Physical Society},
  doi = {10.1103/PhysRevLett.132.226601},
  url = {https://link.aps.org/doi/10.1103/PhysRevLett.132.226601}
}

@article{Bunched2003,
  title = {Scattering of Bunched Fractionally Charged Quasiparticles},
  author = {Chung, Y. C. and Heiblum, M. and Umansky, V.},
  journal = {Phys. Rev. Lett.},
  volume = {91},
  issue = {21},
  pages = {216804},
  numpages = {4},
  year = {2003},
  month = {Nov},
  publisher = {American Physical Society},
  doi = {10.1103/PhysRevLett.91.216804},
  url = {https://link.aps.org/doi/10.1103/PhysRevLett.91.216804}
}

@article{bid2009,
  title = {Shot Noise and Charge at the $2/3$ Composite Fractional Quantum Hall State},
  author = {Bid, Aveek and Ofek, N. and Heiblum, M. and Umansky, V. and Mahalu, D.},
  journal = {Phys. Rev. Lett.},
  volume = {103},
  issue = {23},
  pages = {236802},
  numpages = {4},
  year = {2009},
  month = {Dec},
  publisher = {American Physical Society},
  doi = {10.1103/PhysRevLett.103.236802},
  url = {https://link.aps.org/doi/10.1103/PhysRevLett.103.236802}
}

@article{Gefen_2021,
  title = {Fractional edge reconstruction in integer quantum Hall phases},
  author = {Khanna, Udit and Goldstein, Moshe and Gefen, Yuval},
  journal = {Phys. Rev. B},
  volume = {103},
  issue = {12},
  pages = {L121302},
  numpages = {6},
  year = {2021},
  month = {Mar},
  publisher = {American Physical Society},
  doi = {10.1103/PhysRevB.103.L121302},
  url = {https://link.aps.org/doi/10.1103/PhysRevB.103.L121302}
}

@Article{Biswas2022,
author={Biswas, Sourav
and Bhattacharyya, Rajarshi
and Kundu, Hemanta Kumar
and Das, Ankur
and Heiblum, Moty
and Umansky, Vladimir
and Goldstein, Moshe
and Gefen, Yuval},
title={Shot noise does not always provide the quasiparticle charge},
journal={Nature Physics},
year={2022},
month={Dec},
day={01},
volume={18},
number={12},
pages={1476-1481},
issn={1745-2481},
doi={10.1038/s41567-022-01758-x},
url={https://doi.org/10.1038/s41567-022-01758-x}
}

@Article{Sabo2017,
author={Sabo, Ron
and Gurman, Itamar
and Rosenblatt, Amir
and Lafont, Fabien
and Banitt, Daniel
and Park, Jinhong
and Heiblum, Moty
and Gefen, Yuval
and Umansky, Vladimir
and Mahalu, Diana},
title={Edge reconstruction in fractional quantum Hall states},
journal={Nature Physics},
year={2017},
month={May},
day={01},
volume={13},
number={5},
pages={491-496},
issn={1745-2481},
doi={10.1038/nphys4010},
url={https://doi.org/10.1038/nphys4010}
}

@article{MannaNoise2024,
    author = {Manna, Sourav and Das, Ankur and Gefen, Yuval and Goldstein, Moshe},
    title = {Shot noise as a diagnostic in the   $\nu$=2/3 fractional quantum Hall edge zoo},
    journal = {Low Temperature Physics},
    volume = {50},
    number = {12},
    pages = {1113-1122},
    year = {2024},
    month = {12},
    issn = {1063-777X},
    doi = {10.1063/10.0034344},
    url = {https://doi.org/10.1063/10.0034344},
}

@article{Yuval_incoherent,
  title = {Incoherent transport on the $\ensuremath{\nu}=2/3$ quantum Hall edge},
  author = {Nosiglia, Casey and Park, Jinhong and Rosenow, Bernd and Gefen, Yuval},
  journal = {Phys. Rev. B},
  volume = {98},
  issue = {11},
  pages = {115408},
  numpages = {24},
  year = {2018},
  month = {Sep},
  publisher = {American Physical Society},
  doi = {10.1103/PhysRevB.98.115408},
  url = {https://link.aps.org/doi/10.1103/PhysRevB.98.115408}
}

@article{Saminadayar1997,
  title = {Observation of the $\mathit{e}\mathit{/}3$ Fractionally Charged Laughlin Quasiparticle},
  author = {Saminadayar, L. and Glattli, D. C. and Jin, Y. and Etienne, B.},
  journal = {Phys. Rev. Lett.},
  volume = {79},
  issue = {13},
  pages = {2526--2529},
  numpages = {0},
  year = {1997},
  month = {Sep},
  publisher = {American Physical Society},
  doi = {10.1103/PhysRevLett.79.2526},
  url = {https://link.aps.org/doi/10.1103/PhysRevLett.79.2526}
}

@Article{Picciotto1997,
author={de-Picciotto, R.
and Reznikov, M.
and Heiblum, M.
and Umansky, V.
and Bunin, G.
and Mahalu, D.},
title={Direct observation of a fractional charge},
journal={Nature},
year={1997},
month={Sep},
day={01},
volume={389},
number={6647},
pages={162-164},
issn={1476-4687},
doi={10.1038/38241},
url={https://doi.org/10.1038/38241}
}

@Article{Reznikov1999,
author={Reznikov, M.
and Picciotto, R. de
and Griffiths, T. G.
and Heiblum, M.
and Umansky, V.},
title={Observation of quasiparticles with one-fifth of an electron's charge},
journal={Nature},
year={1999},
month={May},
day={01},
volume={399},
number={6733},
pages={238-241},
issn={1476-4687},
doi={10.1038/20384},
url={https://doi.org/10.1038/20384}
}

@article{Chang2003,
  title = {Chiral Luttinger liquids at the fractional quantum Hall edge},
  author = {Chang, A. M.},
  journal = {Rev. Mod. Phys.},
  volume = {75},
  issue = {4},
  pages = {1449--1505},
  numpages = {0},
  year = {2003},
  month = {Nov},
  publisher = {American Physical Society},
  doi = {10.1103/RevModPhys.75.1449},
  url = {https://link.aps.org/doi/10.1103/RevModPhys.75.1449}
}

@Article{Dolev2008,
author={Dolev, M.
and Heiblum, M.
and Umansky, V.
and Stern, Ady
and Mahalu, D.},
title={Observation of a quarter of an electron charge at the $\nu$ = 5/2 quantum Hall state},
journal={Nature},
year={2008},
month={Apr},
day={01},
volume={452},
number={7189},
pages={829-834},
issn={1476-4687},
doi={10.1038/nature06855},
url={https://doi.org/10.1038/nature06855}
}

@article{Heiblum2020,
author = {Heiblum, Moty and Feldman, D. E.},
title = {Edge probes of topological order},
journal = {International Journal of Modern Physics A},
volume = {35},
number = {18},
pages = {2030009},
year = {2020},
doi = {10.1142/S0217751X20300094},

URL = { 
    
        https://doi.org/10.1142/S0217751X20300094
    
    

}
}
\end{document}